\documentclass[
 aip,
 amsmath,amssymb,
preprint,%
]{revtex4-1}

\usepackage{graphicx}
\usepackage{dcolumn}
\usepackage{bm}

\usepackage[utf8]{inputenc}
\usepackage[T1]{fontenc}
\usepackage{mathptmx}
\usepackage{etoolbox}
\usepackage{multirow}
\usepackage{caption}
\usepackage{subcaption}
\usepackage{setspace}
\usepackage{algorithm}
\usepackage{algpseudocode}

\makeatletter
\def\@email#1#2{%
 \endgroup
 \patchcmd{\titleblock@produce}
  {\frontmatter@RRAPformat}
  {\frontmatter@RRAPformat{\produce@RRAP{*#1\href{mailto:#2}{#2}}}\frontmatter@RRAPformat}
  {}{}
}%
\makeatother
\begin{document}

\preprint{AIP/123-QED}

\title[Hybrid VMAT-3DCRT]{Hybrid VMAT-3DCRT as Breast Treatment Improvement Tool}
\author{Cyril Voyant}
 \altaffiliation[Also at ]{SPE Laboratory, University of Corsica.}
 \email{voyant\_c@univ-corse.fr}
\author{Morgane Pinpin}%
\affiliation{ 
Radiation Unit, Castelluccio Hospital, 20090 Ajaccio
}%
\author{Delphine Leschi}%
\author{Séverine Prapant}%
\author{Françoise Savigny}%
\author{Marie-Aimée Acquaviva}%
\affiliation{ 
Radiation Unit, Castelluccio Hospital, 20090 Ajaccio
}%


\date{\today}

\begin{abstract}
Radiation therapy is an important tool in the treatment of breast cancer and can play a crucial role in improving patient outcomes. For breast cancer, if the technique has been for a long time the use of 3DCRT, clinicians have seen the management evolve greatly in recent years. Field-in-field and IMRT approaches and more recently dynamic arctherapy are increasingly present. All of these approaches are constantly trying to improve tumour coverage and to preserve organs at risk by minimising the doses delivered to them. If arctherapy allows a considerable reduction of high doses received by healthy tissues, no one can deny that it also leads to an increase of low doses in tissues that would not have received any with other techniques. It is with this in mind that we propose a hybrid approach combining the robustness of the 3DCRT approach and the high technicality and efficiency of arctherapy. Statistical tests allow us to draw conclusions about the possibility of using the hybrid approach in certain cases (right breast, BMI > 23, age > 48, target volume > 350cc, etc.). Indeed, depending on the breast laterality and patients morphological characteristics, hybridization may prove to be a therapeutic tool of choice in the management of breast cancer in radiotherapy.
 
\end{abstract}

\maketitle

\section{\label{sec:1}Introduction}

Improving the quality of breast cancer treatment is essential to reduce mortality and increase survival rates, as breast cancer is one of the most common cancers in women of all ages \cite{10.1001/jama.2018.19323}. The age of onset can vary, with a significant number of cases occurring in women over 50 but sometimes in younger women. Early detection and prompt treatment can lead to more positive outcomes, so continued efforts to improve breast cancer treatment and care are essential. Thanks to advances in technology, diagnosis and treatment methods, breast cancer survival rates have improved significantly. However, there is still much room for improvement, and ongoing research, clinical and dosimetric trials are essential to advancing the quality of care.

\subsection{\label{sec:11}Radiation Therapy Role in Breast Cancer}

Radiotherapy is an important tool in the treatment of breast cancer and can play a crucial role in improving patient outcomes \cite{FRANCO2023100556}. One of the main advantages of radiotherapy is its ability to target and destroy cancer cells while minimising damage to healthy tissue \cite{ABDOLLAHI2023100201}. This makes it an effective option for treating breast cancer, particularly when used at an early stage \cite{CHUNG2013959}. 
Intensity modulated radiation therapy (IMRT), volumetric modulated arc therapy (VMAT) and three-dimensional conformal radiation therapy (3DCRT) are all forms of radiation therapy used in the treatment of breast cancer \cite{GLEESON2022264}.
IMRT is known for its ability to deliver high doses of radiation to the tumour while minimising high dose exposure to surrounding healthy tissue \cite{KWA19981}. This increases the accuracy and precision of cancer targeting and helps reduce side effects\cite{White2006}. 
VMAT also delivers high doses of radiation to the tumour while minimising exposure of surrounding healthy tissue\cite{cardona2023} and has the advantage of a faster treatment time \cite{ROSSI202186}. However, it is linked to a frequent increase in low doses in healthy tissues adjacent to the target volumes.
3DCRT is the reference and is a widely available and cost-effective form of radiotherapy, but it is less accurate and precise in targeting cancer than IMRT and VMAT, and results in greater exposure of surrounding healthy tissue \cite{BORGER20071131}.
It is important to note that exposure to high and low doses of radiation\cite{taylor_estimating_2017}, can increase the risk of damage to the heart \cite{CHUNG2013959} and lungs \cite{ROY2021155}. 

\subsection{\label{sec:12}VMAT and Low Doses (<10Gy)}

What is known since the introduction of VMAT in breast cancer, is that it is a fabulous technique but that it induces a considerable increase in the volume of healthy tissue receiving low doses. Many teams are wondering about the possible deleterious radiobiological effects of these doses. 
This exposure can have long-term effects on the lungs \cite{cancers13010022}, including a slight increase in the risk of lung cancer, fibrosis (scarring of the lung tissue) and impaired lung function. Studies have shown that exposure to low doses, can lead to changes in the DNA of lung cells that can result in mutations and an increased risk of lung cancer. It is important to note that the risk of lung damage from low-dose radiation exposure during radiotherapy depends on several factors, including the patient's age, general health and the specifics of the radiotherapy such as dose, fractionation and irradiated volume. Several references support these conclusions, including the papers of \citet{iiii} and \citet{HURKMANS2000145}. It should be noted that some papers suggest fatal lung disease as a function of low dose exposure after radiotherapy \cite{KEFFER2020238}, so it seems important to take these low doses into account and to propose new approaches accordingly. 
The same conclusions could be made by focusing on the effects on the heart as shown in \citet{10.1371/journal.pone.0252552}. Indeed, exposure of the heart to low-dose radiation can increase the risk of long-term cardiovascular effects, such as coronary heart disease and heart failure. The precise dose-response relationship for these effects is not well understood, but the risk is generally thought to increase with dose. 
It is likely that other organs are affected by low doses, but in the absence of literature it is best to apply the precautionary principle and try to minimise the use of techniques that induce low dose exposure to organs at risk. It is important to keep in mind that there is no direct evidence indicating that VMAT is not appropriate for breast cancer treatment, but it is important to use caution when using VMAT for large volumes of tissue that may receive low doses of radiation and to determine the best approach for each patient.

\subsection{\label{sec:13}The Choice of Hybrid Approach (3DCRT \& VMAT)}

The advantage of models combination in engineering and physical sciences is that it allows the strengths of multiple models to be combined to overcome the limitations of a single model. 
Each model brings its own strengths and limitations, and their combination can provide a more complete view of the system and lead to more accurate and robust actions. It is in this perspective that it has been proposing for some years in radiotherapy, a hybridization of 3DCRT and VMAT models\cite{liu_dosimetric_2020}. For breast cancer radiotherapy, a hybrid approach combining 3DCRT and VMAT may offer several advantages \cite{doi_hybrid_2020}. 3DCRT is a traditional and modulation-free technique inducing lower coverage of target nodes and may result in higher radiation doses to organs at risk with a volume of low doses in general, restricted. VMAT, on the other hand, uses intensity modulated arcs to deliver a radiation dose more consistent with target volumes (breast and nodes), reduces outside targets high doses but increases low doses.
By combining these techniques, a hybrid approach can deliver a radiation dose more consistent with target volumes \cite{marina_hennet_retrospective_2022}, reduce doses to normal tissues \citet{ashby_late_2021} and reduce the risk of long-term toxicity \citet{CILLA2021295}. The decision to use a hybrid approach should be selective and reserved for certain patients \cite{chen_dosimetric_2020} and treatments \cite{veronesi_value_2008} with nodes irradiation for example \cite{ROSSI2019117}. An effort must be made on the simplicity, speed \cite{VANDURENKOOPMAN2018332} and quality of proposed treatments related to the best PTV coverage and OAR saving compromise \cite{xu_locoregional_2019}.
The structure of this paper is classic with the next section relating to the presentation of data and planning methods (Section \ref{sec:2}), then a part which will deal with all the results (Section \ref{sec:Results}), before proposing a small paragraph dedicated to the technical feasibility of the hybrid method (Section \ref{sec:Feasability}) and concluding (Section \ref{sec:Conclusion}).

\section{\label{sec:2}Material and Methods}

 The patient sample used and the methodology followed throughout the simulations will be detailed in the Sections \ref{sec:Patient Sample} and \ref{sec:Planning}. Then we will expose the comparison metrics used to rank the three planning methods in Section \ref{sec:metrics}.

 \subsection{Patient Sample}
 \label{sec:Patient Sample}
To address the hybridization contribution concerning free breathing breast cancer treatments with nodal prophylactic irradiation (internal mammary chain, intrerpectoral, level 1-4 axillary, etc.), a cohort treated between 2021 and 2022, was used. This retrospective study includes 30 patients (50\% right breasts and 50\% left one). The descriptive statistics are given in the Table ~\ref{tab:desc}.
For each patient, objective validation criteria had to be defined, as well as volumes related to dosimetric optimization. The modalities for generating these volumes were respected for all patients included in the study. In Table ~\ref{tab:voldef} are indicated the setup margins used for the target volumes. The exhaustive list of all parameters used during this study for set-up, optimisation and validation are listed in Table ~\ref{tab:param_dos}. As an example, the clinical objectives are given, but it should be kept in mind that they do not serve as a reference since other criteria could be used. 

\begin{table}
\centering
\caption{Description of patients enrolled in the study concerning the age, volume of 50Gy target (cc; breast) and Body Mass Index (BMI)}
\begin{tabular}{|c|c|c|c|c|c|c|c|c|c|} 
\toprule
         & \multicolumn{3}{c|}{Breast R \& L  } & \multicolumn{3}{c|}{Breast R} & \multicolumn{3}{c|}{Breast L  }  \\ 
\hline
         & BMI   & Age   & Vol-PTV50          & BMI   & Age   & Vol-PTV50     & BMI   & Age   & Vol-PTV50       \\ 
\hline
Mean     & 25.22 & 59.63 & 663.34             & 25.11 & 55.33 & 657.34        & 25.33 & 63.93 & 669.34          \\ 
\hline
SD       & 4.82  & 17.24 & 426.49             & 5.20  & 18.33 & 470.72        & 4.58  & 15.48 & 393.86          \\ 
\hline
Median   & 24.90 & 61.00 & 500.43             & 24.90 & 54.00 & 521.35        & 24.90 & 62.00 & 455.80          \\ 
\hline
max      & 37.46 & 90.00 & 1910.00            & 37.46 & 90.00 & 1910.00       & 34.60 & 88.00 & 1541.00         \\ 
\hline
min      & 16.30 & 25.00 & 151.90             & 16.30 & 25.00 & 151.90        & 19.00 & 42.00 & 257.60          \\ 
\hline
Kurtosis & 0.61  & -0.71 & 1.25               & 1.85  & -0.70 & 2.33          & -0.49 & -1.10 & 0.25            \\ 
\hline
Skewness & 0.60  & 0.03  & 1.26               & 0.85  & 0.14  & 1.44          & 0.33  & 0.25  & 1.14            \\
\hline
\end{tabular}
\label{tab:desc}
\end{table}

\begin{table}
\caption{Volumes definition ($X^C$ for the complement of the set $X$)}
\centering
\renewcommand{\arraystretch}
{0.6}
\begin{tabular}{|c|c|c|}
\toprule
Volume       & Definition              & Remarks                                                        \\ 
\hline
PTV50        & (CTV50+0.5cm) $\cap$ \ (LungIL)$^C$ & CBCT Reg/chest wall. Use for the Optimization  \\
PTV50-Eval   & PTV50 $\cap$ \ (Ext-0,5cm) & Use for the Validation                                        \\
PTV47        & CTV47+0.5cm             & Use for the Optimization                                       \\
PTV47-Eval   & PTV47 $\cap$ \ (Ext-0.5cm) & Use for Validation                                             \\
Bolus Breast \cite{ROSSI2019266}& (PTV50+0.5cm) $\cap$ \ (LungIL)$^C$   & Density =0.9 during Optimization                               \\
Bolus Nodes  & PTV47                   & Density =0.9 during Optimization           \\
\hline
\end{tabular}
\label{tab:voldef}
\end{table}

\begin{table}
\centering
\caption{Dosimetric parameters definition. Note that PTV50 and PTV47 must imperatively be replaced by PTV50-Eval and PTV47-Eval if we had not simplified the labels for the sake of readability}
\label{tab:param_dos}
\renewcommand{\arraystretch}
{0.6}
\begin{tabular}{|c|c|c|c|c|c|} 
\toprule
\# & Structure                     & Parameter      & Definition                              & Unit    & Clinical Goal  \\ 
\cline{1-6}
1  & \multirow{11}{*}{Target 50Gy} & PTV50-D98~     & Dose related to 98\% of volume          & Gy       &45 \\
2  &                               & PTV50-D2       & Dose related to 2\% of volume           & Gy       & 53.5\\
3  &                               & PTV50-D50      & Dose related to 50\% of volume          & Gy       & 50 \\
4  &                               & PTV50-HI       & Homogeneity Index \cite{patel_plan_2020}                      & unitless  & 0.14\\
5  &                               & Vol-PTV50      & Volume of PTV50                         & cc        & n/a \\
6  &                               & Vol-iso95      & Volume of isodose 95\%                  & cc        & n/a\\
7  &                               & Vol-intersec   & Vol-PTV50 $\cap$ \ Vol-iso95                     & cc     &  n/a  \\
8  &                               & PTV50-CI       & Conformity Index \cite{patel_plan_2020}                       & unitless  & 0.5\\
9  &                               & PTV50-V107     & Volume of 107\% of dose(53,5Gy) & \%       & 1 \\
10 &                               & PTV50-V95      & Volume of 95\% of dose(47,5Gy)  & \%       & 90 \\
11 &                               & PTV50-V98      & Volume of 98\% of dose(49Gy)    & \%       &  80\\ 
\cline{1-6}
12 & \multirow{2}{*}{Tagets 47Gy}  & PTV47-V95      & Volume of 95\% of dose(44,65Gy) & \%       & 95 \\
13 &                               & PTV47-V98      & Volume of 98\% of dose(46,06Gy) & \%       & 80 \\ 
\cline{1-6}
14 & \multirow{4}{*}{Lung ipsolat} & LungIL-Dmean   & Mean Dose                               & Gy       & 15 \\
15 &                               & LungIL-V20     & Volume related to 20Gy                  & \%        & 30 \\
16 &                               & LungIL-V30     & Volume related to 30Gy                  & \%        & 20\\
17 &                               & LungIL-NTCP    & NTCP related to Lyman model             & \%       & 5 \\ 
\cline{1-6}
18 & Lung contralat                & LungCL-Dmean   & Mean Dose                               & Gy       & 5 \\ 
\cline{1-6}
19 & Lungs     (IL $\cup$ CL)                    & LungILCL-V5    & Volume related to 5Gy                   & \%       & 50 \\ 
\cline{1-6}
20 & \multirow{3}{*}{Heart}        & Heart-Dmean~   & Mean Dose                               & Gy       & 5 \\
21 &                               & Heart-V25      & Volume related to 25Gy                  & \%       & 10 \\
22 &                               & AIV-V30        & Volume related to 30Gy (AIV)            & \%       & 30 \\ 
\cline{1-6}
23 & Liver                         & Liver-V5       & Volume related to 5Gy                   & cc       & 100 \\ 
\cline{1-6}
24 & Breast contralat              & BreastCL-Dmean & Mean Dose                               & Gy       & 5 \\ 
\cline{1-6}
25 & Humeral head                  & HH-Dmean       & Mean Dose                               & Gy       & 20 \\ 
\cline{1-6}
26 & spinal cord (+3mm)            & PRVSP-Dmax     & Max Dose                                & Gy       & 20 \\ 
\cline{1-6}
27 & Esophagus                     & Eso-V35~       & Volume related to 35Gy                  & cc       & 5 \\ 
\cline{1-6}
28 & Trachea                       & Trachea-V35    & Volume related to 35Gy                  & cc       & 5 \\
\hline
\end{tabular}
\end{table}

 \subsection{Treatments Planning}
\label{sec:Planning}
The plan followed in this study is relatively simple, we had, for each patient established three dosimetries (field-in-field based 3DCRT, VMAT and Hybrid) and several dosimetric parameters (listed in Table \ref{tab:param_dos}) were collected. IMRT was not included in this study since the results are close to those obtained with the field-in-field technique. The sample (30 patients) will allow to use different statistical metrics (defined in the following subsection) to conclude whether model combination is useful. 
Some rules were followed throughout the simulations. All contours (CTV \& OAR) were equitably distributed between three physicians and plannings between two physicists. The study guideline were :
\begin{singlespace}
\begin{itemize}
  \item Isocentre positioned between supraclavicular nodes and the mammary gland in cranio-caudal direction and close to the center line in the internal-external one, so that realise CBCT is technically possible;
  \item Same isocentre for VMAT, 3DCRT and hybrid;
  \item Each trial (VMAT, Hybrid and 3DCRT) is done without comparison with the previous ones so as not to be tempted to enter into a logic of "fine-tuning" which would bias the study;
  \item Dynamic leaf gap of 1cm for all plans using arctherapy in order to improve plans quality control; 
   \item VMAT plans made using the virtual bolus technique \cite{https://doi.org/10.1002/acm2.12398};
  \item 3DCRT plans use different energies, wedge, fields-in-fields, etc. anything that allows to establish an acceptable output.
\end{itemize}
\end{singlespace}

We insisted on proposing the most objective study possible by scrupulously respecting the rules set out above, as well as the definition of the beams that are identical for all patients and respects the characteristics set out in Table ~\ref{tab:Beam}. The characteristics of arcs used during VMAT and hybrid planning are detailed in Table ~\ref{tab:arc}. Simulations concern the Treatment Planning System Pinnacle (V16.4), two Elekta linacs (Synergy with MLC Agility) and free Breathing slow-CT acquisitions (BigBore Philips). Concerning the run$\#$1 of arcs optimization, the same process (Table ~\ref{tab:opti}) was applied for all patients. During next runs, planners modified PTV and OAR optimization criteria as desired \cite{MAHE2016S4}. To eliminate hot spots, most of the time, several passes (between 2 and 4) were necessary according to Algorithm \ref{algo1} where No Man's Land (NML) is defined from the isodose 107\% ($\lor$ and $\land$ are respectively logical OR and AND). 

\begin{algorithm} [H]
\caption{Hot Spots Removal}
\label{algo1}
\begin{algorithmic} [1]
\Require $ (D_{max}>1.1\times 50Gy) ~\lor ~(V107\%>1\%) $
\Ensure $ \mathrm{Other ~Clinical ~Goals ~Achieved}$
\State $NML \leftarrow iso107\%$
\Repeat $~NML\leftarrow NML(+1)$
\Until {$(D_{max}<55Gy)$ $~\land$ ~$(V53.5Gy<1\%)$}
\State Density(Bolus Breast $\cup$ Bolus Nodes) $\leftarrow$ Override ($=0.9$)
\State Constrain Optimization with $D_{max}(NML)<53.5Gy$ (Priority$~=200$)
\State Density(Bolus Breast $\cup$ Bolus Nodes) $\leftarrow$ Reverse Override
\State Dose Computing
\State \Return Finalized Plan
\end{algorithmic}
\end{algorithm}

\begin{table}
\caption{Beams definition}
\centering
\renewcommand{\arraystretch}
{0.6}
\begin{tabular}{|c|c|c|c|c|c|c|c|} 
\toprule

Method                  & Prescription                                                          & Beams             & Nature                 & Angle (R) & Rotation & Angle (L) & Colli           \\ 
\hline
\multirow{3}{*}{3DCRT}  & \multirow{2}{*}{Breast+IMN\footnote{internal mammary nodal region}} & TGint             & FiF (X6/18)           & 55                 &   n/a       & 305                & 0               \\
                        &                                                                       & TGext             & Wedged (X6/18)        & 233                &    n/a      & 127                & 90/270       \\
                        & Other                                                                 & 3 beams & \{Wed.+FiF\} (X6/18) & 250/275/30         &     n/a     & 110/85/330         & 0/90/270  \\ 
\hline
\multirow{3}{*}{VMAT}   & \multirow{3}{*}{Breast+Nodes}                                     & Arc1-1            & VMAT (X6)              & 200/250            & CC       & 290/350            & 0               \\
                        &                                                                       & Arc1-2            & VMAT (X6)              & 10/70              & CC       & 110/160            & 0               \\
                        &                                                                       & Arc1-3            & VMAT (X6)              & 70/200             & CCW      & 160/290            & 90              \\ 
\hline
\multirow{3}{*}{Hybrid} & \multirow{2}{*}{Breast+IMN}                                           & TGint             & Open (X6)              & 55                 &   n/a       & 305                & 0               \\
                        &                                                                       & TGext             & Wedged (X6)            & 233                &   n/a       & 127                & 90/270       \\
                        & Breast+Others                                                     & Arc               & VMAT (X6)              & 200/70             & CCW      & 160/290            & 0               \\
\hline
\end{tabular}
\label{tab:Beam}
\end{table}

\begin{table}
\centering
\caption{Arcs definition (Pinnacle 16.4)}
\renewcommand{\arraystretch}
{0.6}
\begin{tabular}{|c|c|c|c|c|c|c|c|c|c|c|c|} 
\toprule
\multirow{2}{*}{Method } & \multirow{2}{*}{Beam} & Max
  Deliv & Gant
                & Seg
             & Lea           & Conv             & Cont
             &       MU                  & Dose
  fall      & Targ
          & Targ
  Edge        \\
                         &                         & Time (s)    & Spac & Area           & Mot         & Cycles              & Mod                     & 
  level              & Off              & ~Falloff     & ~Weight             \\ 
\hline
\multirow{3}{*}{VMAT}    & Arc1-1                  & 50          & \multirow{4}{*}{2° }       & \multirow{4}{*}{15cc} & \multirow{4}{*}{0,4cm/°} & \multirow{4}{*}{6/2} & \multirow{4}{*}{Med} & \multirow{4}{*}{Med} & \multirow{4}{*}{25} & \multirow{4}{*}{40} & \multirow{4}{*}{2}  \\
                         & Arc1-2                  & 50          &                           &                     &                      &                      &                         &                         &                     &                     &                     \\
                         & Arc1-3                  & 100         &                           &                     &                      &                      &                         &                         &                     &                     &                     \\
Hybrid                   & Arc                     & 100         &                           &                     &                      &                      &                         &                         &                     &                     &                     \\
\hline
\end{tabular}
\label{tab:arc}
\end{table}

\begin{table}
\centering
\caption{Optimization Initialyzing (run \# 1) }
\begin{tabular}{|l|l|l|l|} 
\toprule
Volume                 & Type          & Target & Priority   \\ 
\hline
\multirow{2}{*}{PTV50} & D95           & 47.5   & Defaut     \\ 
\cline{2-4}
                       & Uniforme Dose & 50.5   & Defaut     \\ 
\hline
\multirow{2}{*}{PTV47} & D95           & 44.65  & Defaut     \\ 
\cline{2-4}
                       & Uniforme Dose & 47.5   & Defaut     \\ 
\hline
PRVSP                  & Dmax          & 33     & Very High  \\ 
\hline
Lung ipso              & Max EUD       & 12     & Medium     \\ 
\hline
Lung contra            & Max EUD       & 5      & Medium     \\ 
\hline
Heart                  & Max EUD       & 5      & Medium     \\ 
\hline
AIV                    & Max EUD       & 30     & Medium     \\ 
\hline
Breast contra          & Max EUD       & 5      & Medium     \\ 
\hline
HH                     & Max Dose      & 43     & Medium     \\ 
\hline
PRVmed  \footnote{Defined by (larynx $\cup$ \ esophagus $\cup$ \ thyroid)+0.3cm}               & Max EUD       & 30     & Medium     \\
\hline
\end{tabular}
\label{tab:opti}
\end{table}

\subsection{Comparison Metrics}
\label{sec:metrics}
Dosimetric comparison was made by the use of classical statistical tools. Knowledgeable reader will notice in the following that the normality of the distributions is not clearly established, imposing non-parametric hypothesis tests. Wilcoxon rank-sum and ANOVA tests will be proposed in order to test significant differences between distributions. The use of  coefficients of determination and correlation will complete the analysis, by making it possible to estimate the statistical link between dosimetric quantities. In the last part of the results, analyzes through receiver operating characteristic (ROC) curves will be proposed, testing if ``a priori'' factors (age, volume of 50Gy target and BMI) could make it possible to highlight thresholds below which the clinical objectives (Table ~\ref{tab:param_dos}) are statistically achieved. The area of ROC curves (AROC)  also referred to as area under curve (AUC) will be used to this task. It is important to notice that this test is closely related to the Wilcoxon one. 

\section{\label{sec:Results}Results}
Several tools were used to compare the three planning methods presented above. We will start with probability distributions coupled with an ANOVA significance test (Part \ref{sec:Visual}), then  means comparison non-parametric test (Part \ref{sec:Mean}), followed by coefficients of determination and a visual comparative approach (Part \ref{sec:Determination}) preceding a comparison based on the rank correlation coefficient (Part \ref{sec:Spearman}). This is followed by a study of ROC curves (Part \ref{sec:ROC}). Findings will be discussed in each part and then the main ones will be included in the conclusion. All the data and codes related to this study are available in \url{https://github.com/cyrilvoyant/Hybrid}.

\subsection{Probability Density Function}
\label{sec:Visual}
\begin{figure}
 \caption{Violin plots of all parameters for the three planning methods and associated nonparametric one-way ANOVA on ranks test (pvalue$<0.05$ induces that the 3 distributions are not related to the same population)}
     \centering
     \begin{subfigure}[b]{0.49\textwidth}
         \centering
         \includegraphics[width=\textwidth]{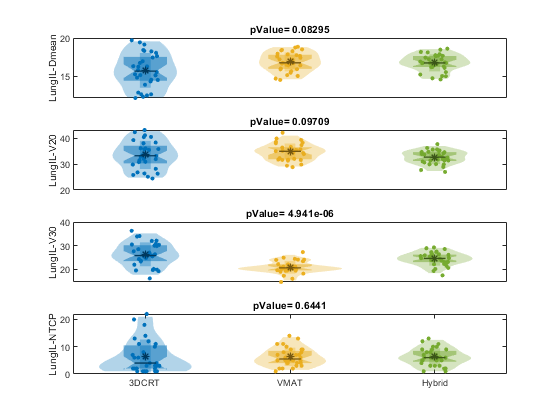}
         \end{subfigure}
           \begin{subfigure}[b]{0.49\textwidth}
         \centering
         \includegraphics[width=\textwidth]{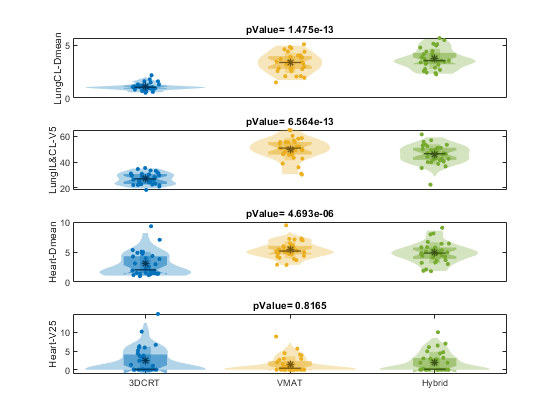}
         \end{subfigure}
           \begin{subfigure}[b]{0.49\textwidth}
         \centering
         \includegraphics[width=\textwidth]{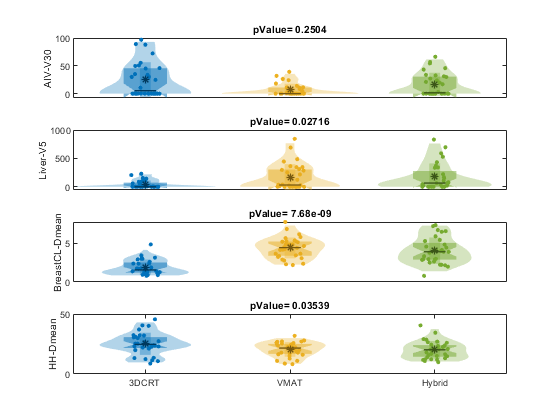}
         \end{subfigure}
         \begin{subfigure}[b]{0.49\textwidth}
         \centering
         \includegraphics[width=\textwidth]{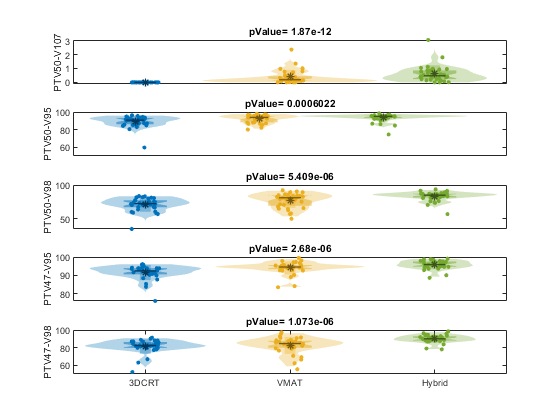}
         \end{subfigure}
           \begin{subfigure}[b]{0.49\textwidth}
         \centering
         \includegraphics[width=\textwidth]{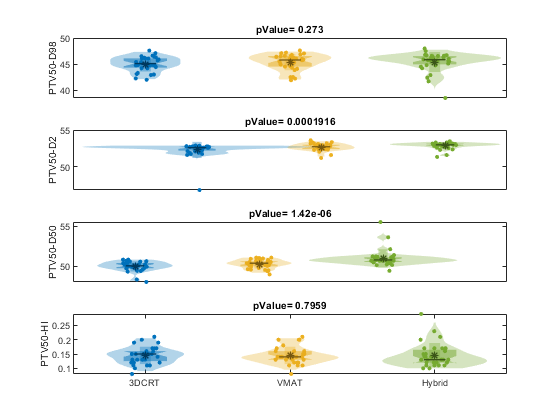}
         \end{subfigure}
           \begin{subfigure}[b]{0.49\textwidth}
         \centering
         \includegraphics[width=\textwidth]{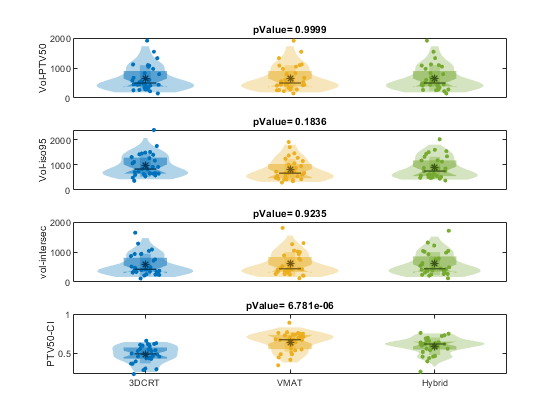}
         \end{subfigure}
         \begin{subfigure}[b]{0.49\textwidth}
         \centering
         \includegraphics[width=\textwidth]{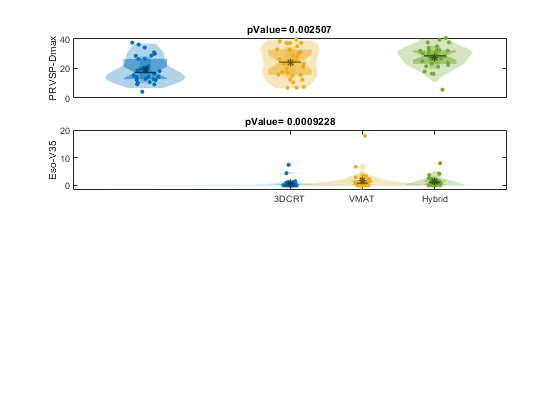}
         \end{subfigure}
\label{fig:violin}
 \end{figure}

In Fig ~\ref{fig:violin} are represented the probability distributions (according to violin plots built with 30 patients) of parameters previously presented in Table \ref{tab:param_dos}. The three studied methods (3DCRT, VMAT and Hybrid) are considered. Results of non parametric one-way ANOVA (pvalue) is shown in order to make easy the interpretation of the distributions comparison. The first important element that is visible is that the distributions are (mostly) not Gaussian, which leads to an inconsistent normal hypothesis. This is why we have favoured non-parametric statistical tools. Among the graphs that stand out, it is worth noting that for lungIL, only the V30 shows a significant difference between the three distribution types. The VMAT being the method showing the best results. For LungCL and LungILCL, there is also an extremely significant difference highlighting the quality of the 3DCRT dosimetry. For the other parameters, there is nothing really conclusive and even if it is clear that the distributions are not from the same population (pvalue < 0.5) the averages are relatively close and further studies are needed to conclude. In the following, we will separate three cases, all patients (n=30), those who had treatment on the right breast (n=15) and those on the left one (n=15).

\subsection{Mean Comparison and Nonparametric Test}
\label{sec:Mean}
\begin{figure}
 \caption{Isodoses related to the three treatments (from left to right : 3DCRT, VMAT and Hybrid) }
     \centering
        
         \includegraphics[width=\textwidth]{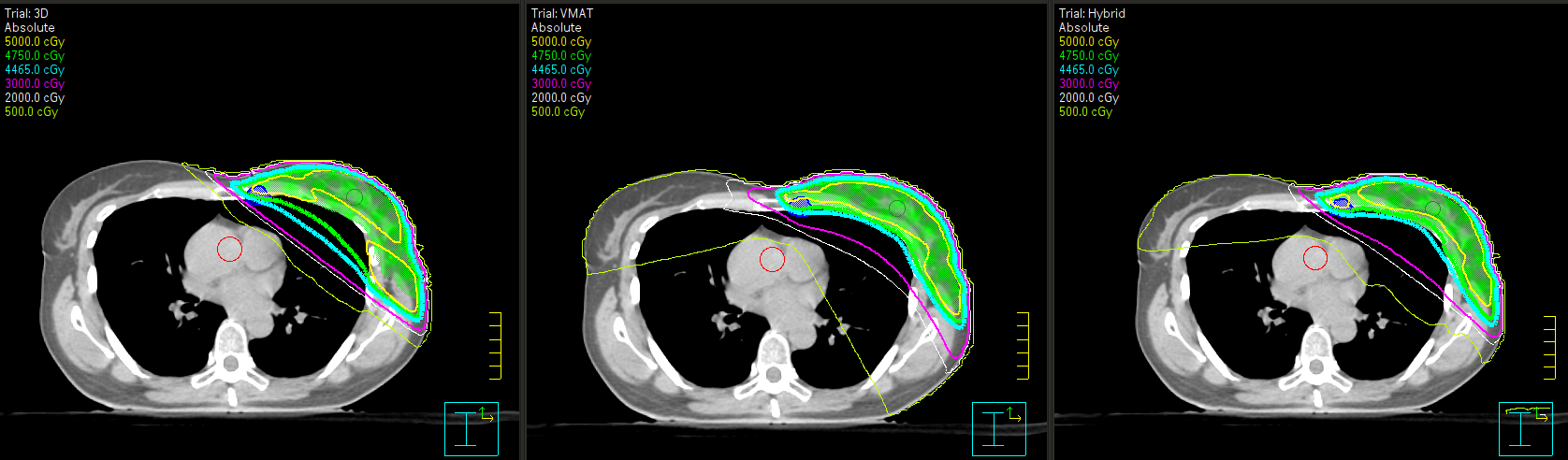}
         \label{fig:isodose}
         \end{figure}

The means of all parameters for all treatment techniques used are available in the Table ~\ref{tab:mean}. Both bold and italic values allow rank the three planning methods for each dosimetric variables (13 concerning the target volumes and 15 considering the organes at risk). At first sight, a trend seems to be emerging, if hybrid appears to be the method of choice for target volumes, 3DCRT is for organs at risk. On closer inspection, the differences are often not large, which encourages the use of hypothesis testing in the future. For ease but above all in order to make the comparison objective, we have decided to propose a non-parametric test on pairwise comparisons. This corresponds to three  scenarios observed in Table ~\ref{tab:pvalue}. First, considering 3DCRT and VMAT, this last is the best way considering PTV50 and PTV47 for all the three samples. Mean dose related to LungIL and V5 of LungILCL are undeniably minimised by the 3DCRT method as shown in Figure \ref{fig:isodose}.

\begin{table}
\centering
\caption{Mean Values (bold for best results and italic for worst)}
\renewcommand{\arraystretch}
{0.6}
\begin{tabular}{|c|ccc|ccc|ccc|} 
\toprule
                 & \multicolumn{3}{c|}{Breast
  RL}                              & \multicolumn{3}{c|}{Breast
  R}                                & \multicolumn{3}{c|}{Breast
  L}                                \\
Parameters       & 3DCRT            & VMAT             & Hybrid           & 3DCRT             & VMAT             & Hybrid           & 3DCRT            & VMAT             & Hybrid            \\ 
\hline
PTV50-D98      & \textit{ 45.02} &  45.38          & \textbf{ 45.4} & \textit{ 45.21}  &  45.93          & \textbf{ 46.03} & \textbf{ 44.82} & \textbf{ 44.82} & \textit{ 44.76}  \\
PTV50-D2       & \textit{ 52.3} &  52.68          & \textbf{ 52.88} & \textit{ 52.48}  &  52.64          & \textbf{ 52.78} & \textit{ 52.11} &  52.72          & \textbf{ 52.98}  \\
PTV50-D50      & \textit{ 50.01} &  50.29          & \textbf{ 51.04} & \textit{ 49.95}  &  50.29          & \textbf{ 51.26} & \textit{ 50.07} &  50.29          & \textbf{ 50.81}  \\
PTV50-HI       & \textbf{ 0.145} & \textbf{ 0.145} & \textit{ 0.146} & \textit{ 0.1447}  &  0.1327          & \textbf{ 0.13} & \textbf{ 0.1453} &  0.1573          & \textit{ 0.1627}  \\
Vol-PTV50      &  663.3          &  663.3          &  663.3          &  657.3           &  657.3          &  657.3          &  669.3          &  669.3          &  669.3           \\
Vol-iso95      &  980.1          &  804.9          &  891.5          & 1004                &  812.1          &  910.8          &  956.6          &  797.7          & \textit{ 872.2}  \\
Vol-intersec   &  576.7          & 603                &  609.2          & 574                 &  604.6          &  611.1          &  579.3          &  601.4          &  607.3           \\
PTV50-CI       & \textit{ 0.487} & \textbf{ 0.645} &  0.5917          & \textit{ 0.471}  & \textbf{ 0.629} &  0.578          & \textit{ 0.504} & \textbf{ 0.661} &  0.6053           \\
PTV50-V107     & \textbf{0}         &  0.3923          & \textit{ 0.612} & \textbf{0}          &  0.2967          & \textit{ 0.6867} & \textbf{0}         &  0.488          & \textit{ 0.5373}  \\
PTV50-V95      & \textit{ 88.94} &  92.24          & \textbf{ 93.33} & \textit{ 89.4}  &  93.12          & \textbf{ 94.4} & \textit{ 88.48} &  91.36          & \textbf{ 92.27}  \\
PTV50-V98      & \textit{ 70.91} &  77.04          & \textbf{ 83.07} & \textit{ 70.28}  &  77.6          & \textbf{ 84.14} & \textit{ 71.53} &  76.49          & \textbf{82}          \\
PTV47-V95      & \textit{ 91.51} &  94.03          & \textbf{ 95.73} & \textit{ 90.46}  &  94.64          & \textbf{ 95.39} & \textit{ 92.57} &  93.41          & \textbf{ 96.06}  \\
PTV47-V98      & \textit{ 81.19} &  82.49          & \textbf{ 90.06} & \textit{ 78.88}  &  82.91          & \textbf{ 89.22} & \textit{ 83.5} &  82.06          & \textbf{ 90.91}  \\ 
\hline
LungIL-Dmean   & \textbf{ 15.74} & \textit{ 16.86} &  16.73          & \textbf{ 16.5}  & \textit{ 16.93} &  16.79          & \textbf{ 14.98} & \textit{ 16.79} &  16.67           \\
LungIL-V20     &  33.51          & \textit{ 34.69} & \textbf{ 32.63} & \textit{ 35.32}  &  34.15          & \textbf{ 33.36} & \textbf{ 31.7} & \textit{ 35.23} &  31.89           \\
LungIL-V30     & \textit{ 26.31} & \textbf{ 20.85} &  24.36          & \textit{ 27.81}  & \textbf{ 20.58} &  24.71          & \textit{ 24.8} & \textbf{ 21.11} &  24.01           \\
LungIL-NTCP    & \textit{ 6.6} &  6.3          & \textbf{ 6.267} & \textit{ 8.333}  & \textbf{ 6.667} &  6.8          & \textbf{ 4.867} & \textit{ 5.933} &  5.733           \\
LungCL-Dmean   & \textbf{ 1.075} &  3.357          & \textit{ 3.736} & \textbf{ 1.067}  &  3.274          & \textit{ 3.981} & \textbf{ 1.082} &  3.44          & \textit{ 3.49}  \\
LungILCL-V5    & \textbf{ 26.94} & \textit{ 50.02} &  45.99          & \textbf{ 30.67}  & \textit{ 53.97} &  50.51          & \textbf{ 23.21} & \textit{ 46.08} &  41.47           \\
Heart-Dmean    & \textbf{ 3.004} & \textit{ 5.378} &  4.867          & \textbf{ 1.468}  & \textit{ 4.501} &  3.988          & \textbf{ 4.541} & \textit{ 6.255} &  5.746           \\
Heart-V25      & \textit{ 2.502} & \textbf{ 1.428} &  2.049          & \textbf{0.009} & \textit{ 0.052} & 0.019          & \textit{ 4.995} & \textbf{ 2.803} &  4.079           \\
AIV-V30        & \textit{ 25.25} & \textbf{ 7.075} &  16.88          & \textbf{0}          & \textbf{0}         & \textbf{0}          & \textit{ 50.49} & \textbf{ 14.15} &  33.75           \\
Liver-V5       & \textbf{ 43.16} & 172                & \textit{ 178.8} & \textbf{ 85.74}  &  310.9          & \textit{ 317.5} & \textbf{ 0.576} &  33.14          & \textit{ 40.15}  \\
BreastCL-Dmean & \textbf{ 1.918} & \textit{ 4.405} &  4.047          & \textbf{ 2.482}  & \textit{ 4.448} &  4.378          & \textbf{ 1.355} & \textit{ 4.362} &  3.716           \\
HH-Dmean       & \textit{ 25.16} &  20.46          & \textbf{ 20.1} & \textit{ 24.99}  &  19.59          & \textbf{ 18.75} & \textit{ 25.34} & \textbf{ 21.33} &  21.46           \\
PRVSP-Dmax     & \textbf{ 19.21} &  24.01          & \textit{ 27.62} & \textbf{ 24.14}  &  28.62          & \textit{ 30.45} & \textbf{ 14.28} &  19.41          & \textit{ 24.79}  \\
Eso-V35        & \textbf{ 0.583} & \textit{ 1.728} &  1.471          & \textbf{ 0.206}  &  0.888          & \textit{ 1.229} & \textbf{ 0.959} & \textit{ 2.568} &  1.713           \\
Trachea-V35    & \textbf{ 2.468} & \textit{ 3.545} &  3.32           & \textbf{ 3.056}  &  4.227          & \textit{ 4.673} & \textbf{ 1.88} & \textit{ 2.862} &  1.967            \\
\cline{1-10}
\end{tabular}
\label{tab:mean}
\end{table}

\begin{table}
\centering
\caption{Significant parameters and associated p-value related to Wilcoxon rank-sum nonparametric test}
\renewcommand{\arraystretch}
{0.6}
\begin{tabular}{|l|ll|ll|ll|l} 
\toprule
\multicolumn{1}{|c|}{}                                      & \multicolumn{2}{c|}{\textbf{Breast R\&L }} & \multicolumn{2}{|c|}{\textbf{Breast R }} & \multicolumn{2}{c|}{\textbf{Breast L }} & \multicolumn{1}{c}{}  \\ 
\cline{1-7}
\multicolumn{1}{|c|}{\multirow{14}{*}{\textbf{3D Vs VMAT}}} & PTV50-CI       & p=1.0e-05               & PTV50-CI       & p=0.004             & PTV50-CI       & p=0.001            &                       \\
\multicolumn{1}{|c|}{}                                      & PTV50-V107     & p=1.9e-10               & PTV50-V107     & p=2.7e-05            & PTV50-V107     & p=2.5e-06             &                       \\
\multicolumn{1}{|c|}{}                                      & PTV50-V95      & p=0.008                & PTV50-V95      & p=0.0114               & LungIL-Dmean   & p=0.018              &                       \\
\multicolumn{1}{|c|}{}                                      & PTV50-V98      & p=0.014                 & PTV50-V98      & p=0.036              & LungIL-V20     & p=0.040              &                       \\
\multicolumn{1}{|c|}{}                                      & PTV47-V95      & p=0.001                  & PTV47-V95      & p=0.001            & LungCL-Dmean   & p=4.5e-06            &                       \\
\multicolumn{1}{|c|}{}                                      & LungIL-Dmean   & p=0.039                 & LungIL-V30     & p=0.001            & LungILCL-V5    & p=3.3e-06            &                       \\
\multicolumn{1}{|c|}{}                                      & LungIL-V30     & p=3.5e-05               & LungCL-Dmean   & p=5.0e-06            & Heart-Dmean    & p=0.003             &                       \\
\multicolumn{1}{|c|}{}                                      & LungCL-Dmean   & p=7.0e-11               & LungILCL-V5    & p=3.3e-06            & AIV-V30        & p=9.6e-05            &                       \\
\multicolumn{1}{|c|}{}                                      & LungILCL-V5    & p=1.4e-10               & Heart-Dmean    & p=3.3e-06            & BreastCL-Dmean & p=4.1e-06            &                       \\
\multicolumn{1}{|c|}{}                                      & Heart-Dmean    & p=5.2e-06               & Liver-V5       & p=0.005             &                & ~                      &                       \\
\multicolumn{1}{|c|}{}                                      & Liver-V5       & p=0.029                 & BreastCL-Dmean & p=0.001            &                & ~                      &                       \\
\multicolumn{1}{|c|}{}                                      & BreastCL-Dmean & p=1.5e-08               & Eso-V35        & p=0.013              &                & ~                      &                       \\
\multicolumn{1}{|c|}{}                                      & HH-Dmean       & p=0.033                 & ~              & ~                      &                & ~                      &                       \\
\multicolumn{1}{|c|}{}                                      & Eso-V35        & p=0.002                 & ~              & ~                      & ~              & ~                      &                       \\ 
\cline{1-7}
\multirow{16}{*}{\textbf{\; 3D Vs Hyb}}                    & PTV50-D2       & p=1.2e-05               & PTV50-D2       & p=0.011              & PTV50-D2       & p=0.001            &                       \\
                                                           & PTV50-D50      & p=8.3e-07               & PTV50-D50      & p=0.001            & PTV50-D50      & p=0.001             &                       \\
                                                           & PTV50-CI       & p=0.001               & PTV50-CI       & p=0.006             & PTV50-CI       & p=0.043              &                       \\
                                                           & PTV50-V107     & p=1.6e-11               & PTV50-V107     & p=8.6e-06            & PTV50-V107     & p=6.8e-07            &                       \\
                                                           & PTV50-V95      & p=0.001               & PTV50-V95      & p=0.001             & PTV50-V95      & p=0.027              &                       \\
                                                           & PTV50-V98      & p=7.0e-07                & PTV50-V98      & p=3.3e-05            & PTV50-V98      & p=0.004             &                       \\
                                                           & PTV47-V95      & p=2.3e-06               & PTV47-V95      & p=0.001            & PTV47-V95      & p=0.001             &                       \\
                                                           & PTV47-V98      & p=4.6e-07               & PTV47-V98      & p=0.001            & PTV47-V98      & p=0.001            &                       \\
                                                           & LungCL-Dmean   & p=3.0e-11               & LungIL-V30     & p=0.025               & LungIL-Dmean   & p=0.016              &                       \\
                                                           & LungILCL-V5    & p=2.8e-10                & LungCL-Dmean   & p=3.3e-06            & LungCL-Dmean   & p=3.3e-06            &                       \\
                                                           & Heart-Dmean    & p=0.001                & LungILCL-V5    & p=3.3e-06            & LungILCL-V5    & p=1.6e-05            &                       \\
                                                           & Liver-V5       & p=0.015                 & Heart-Dmean    & p=7.4e-06            & Heart-Dmean    & p=0.040              &                       \\
                                                           & BreastCL-Dmean & p=2.3e-06               & Liver-V5       & p=0.008             & Liver-V5       & p=0.048              &                       \\
                                                           & HH-Dmean       & p=0.023                 & BreastCL-Dmean & p=0.001            & BreastCL-Dmean & p=8.8e-05            &                       \\
                                                           & PRVSP-Dmax     & p=0.000               & PRVSP-Dmax     & p=0.038              & PRVSP-Dmax     & p=0.001            &                       \\
                                                           & Eso-V35        & p=0.001               & Eso-V35        & p=0.001             & ~              & ~                      &                       \\ 
\cline{1-7}
\multirow{8}{*}{\textbf{VMAT Vs Hyb}}                   & PTV50-D50      & p=0.001               & PTV50-D50      & p=0.003             & PTV50-D50      & p=0.037               &                       \\
                                                           & PTV50-CI       & p=0.040                 & PTV47-V98      & p=0.007              & PTV47-V95      & p=0.021              &                       \\
                                                           & PTV50-V107     & p=0.033                 & LungIL-V30     & p=0.001             & PTV47-V98      & p=0.005             &                       \\
                                                           & PTV50-V98      & p=0.019                 & ~              & ~                      & LungIL-V20     & p=0.005             &                       \\
                                                           & PTV47-V95      & p=0.016                 & ~              & ~                      & LungIL-V30     & p=0.011               &                       \\
                                                           & PTV47-V98      & p=0.001               & ~              & ~                      & AIV-V30        & p=0.001             &                       \\
                                                           & LungIL-V20     & p=0.014                 & ~              & ~                      & PRVSP-Dmax     & p=0.038              &                       \\
                                                           & LungIL-V30     & p=2.9e-05               & ~              & ~                      & ~              & ~                      &                       \\
\cline{1-7}
\end{tabular}
\label{tab:pvalue}
\end{table}

\subsection{Determination Coefficient Study}
\label{sec:Determination}
\begin{figure}
 \caption{VMAT and Hybrid comparison for parameters where the determination coefficient ($R^2$) are lower than 0.8. Circles for right breasts and crosses for left one, the colored band corresponds to the non-significance area ($\alpha = 0.05$) so all points outside this band are significant. }
     \centering
     \begin{subfigure}[b]{0.24\textwidth}
         \centering
         \includegraphics[width=\textwidth]{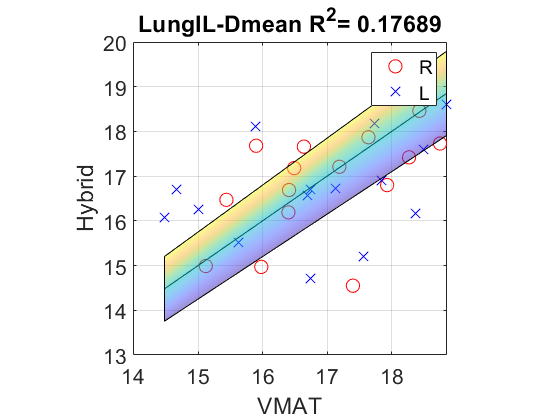}
         \end{subfigure}
           \begin{subfigure}[b]{0.24\textwidth}
         \centering
         \includegraphics[width=\textwidth]{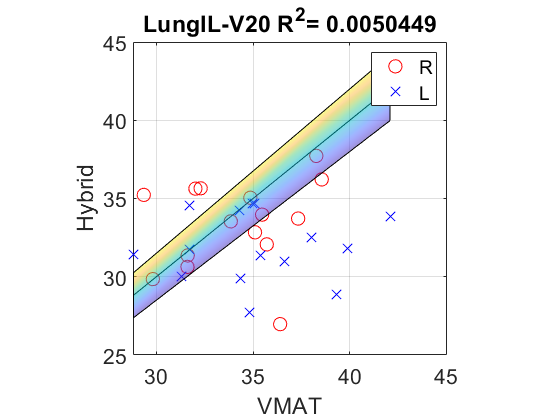}
         \end{subfigure}
           \begin{subfigure}[b]{0.24\textwidth}
         \centering
         \includegraphics[width=\textwidth]{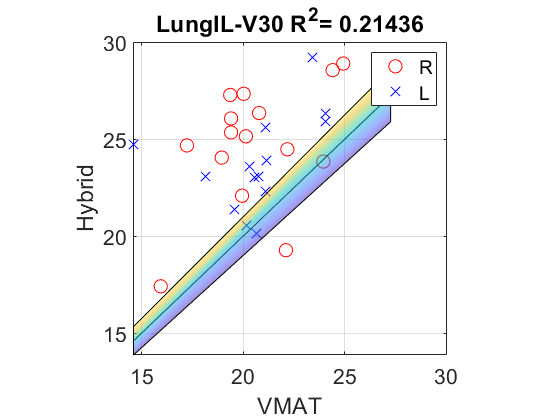}
         \end{subfigure}
           \begin{subfigure}[b]{0.24\textwidth}
         \centering
         \includegraphics[width=\textwidth]{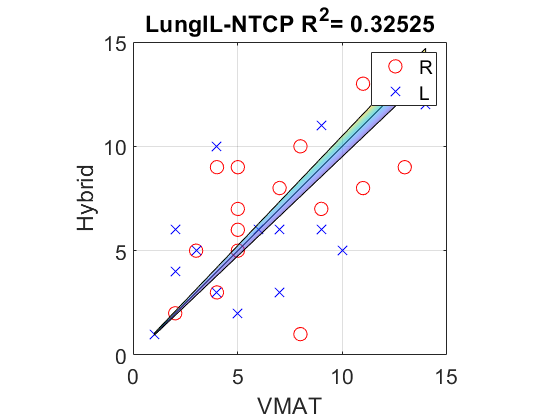}
         \end{subfigure}
              \begin{subfigure}[b]{0.24\textwidth}
         \centering
         \includegraphics[width=\textwidth]{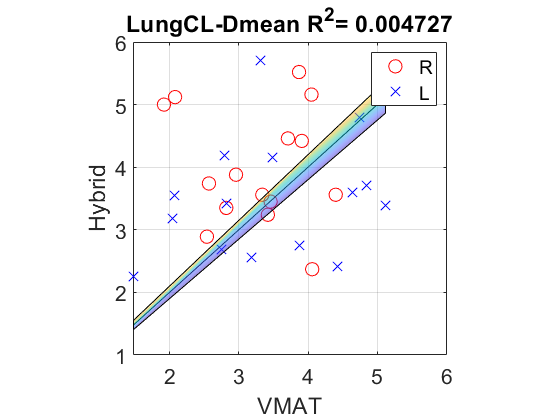}
         \end{subfigure}
           \begin{subfigure}[b]{0.24\textwidth}
         \centering
         \includegraphics[width=\textwidth]{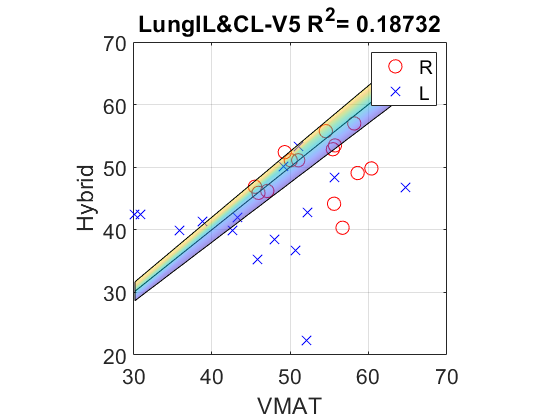}
         \end{subfigure}
           \begin{subfigure}[b]{0.24\textwidth}
         \centering
         \includegraphics[width=\textwidth]{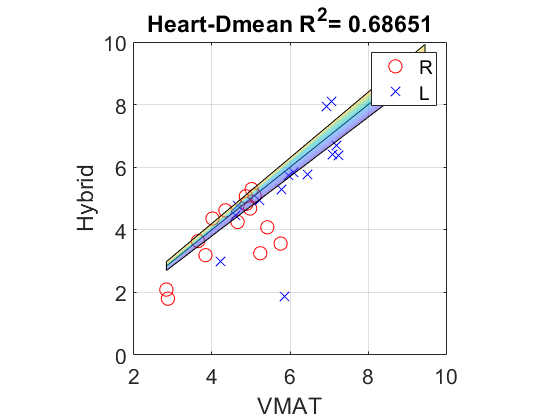}
         \end{subfigure}
           \begin{subfigure}[b]{0.24\textwidth}
         \centering
         \includegraphics[width=\textwidth]{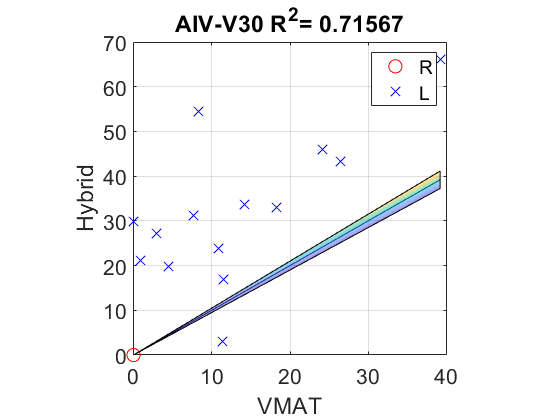}
         \end{subfigure}
         \begin{subfigure}[b]{0.24\textwidth}
         \centering
         \includegraphics[width=\textwidth]{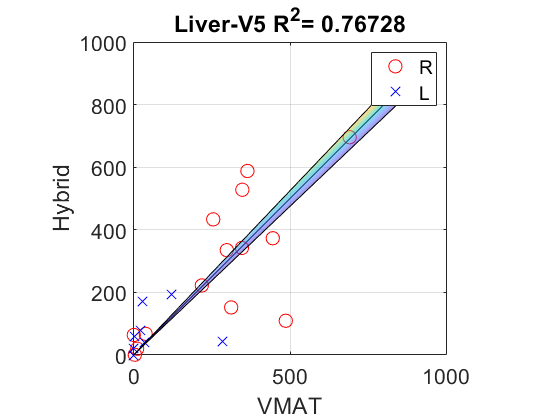}
         \end{subfigure}
              \begin{subfigure}[b]{0.24\textwidth}
         \centering
         \includegraphics[width=\textwidth]{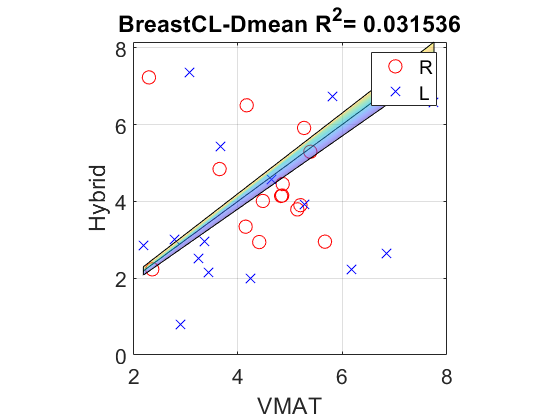}
         \end{subfigure}
           \begin{subfigure}[b]{0.24\textwidth}
         \centering
         \includegraphics[width=\textwidth]{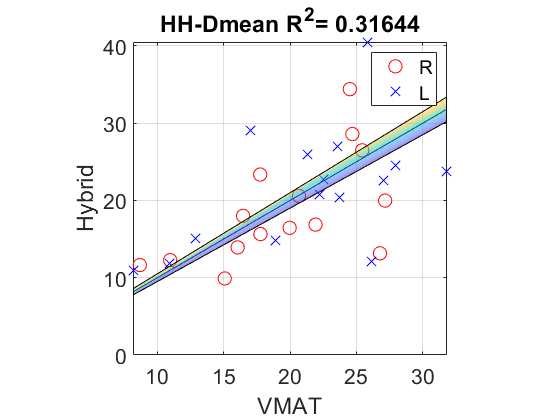}
         \end{subfigure}
           \begin{subfigure}[b]{0.24\textwidth}
         \centering
         \includegraphics[width=\textwidth]{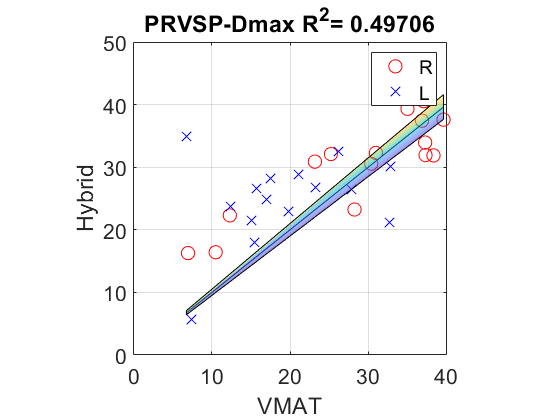}
         \end{subfigure}
           \begin{subfigure}[b]{0.24\textwidth}
         \centering
         \includegraphics[width=\textwidth]{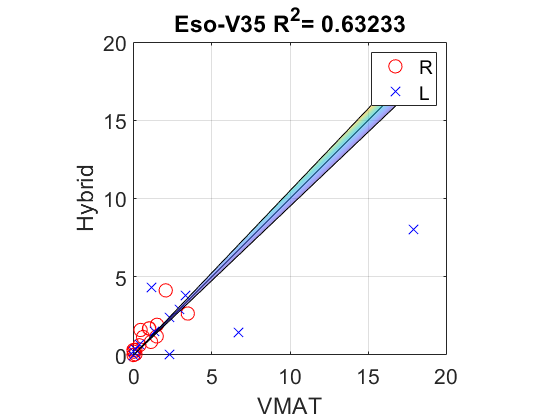}
         \end{subfigure}
              \begin{subfigure}[b]{0.24\textwidth}
         \centering
         \includegraphics[width=\textwidth]{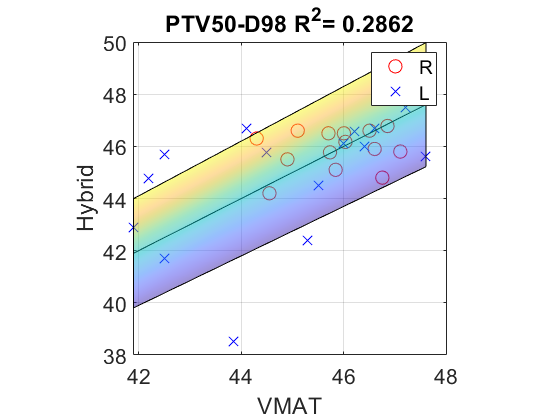}
         \end{subfigure}
           \begin{subfigure}[b]{0.24\textwidth}
         \centering
         \includegraphics[width=\textwidth]{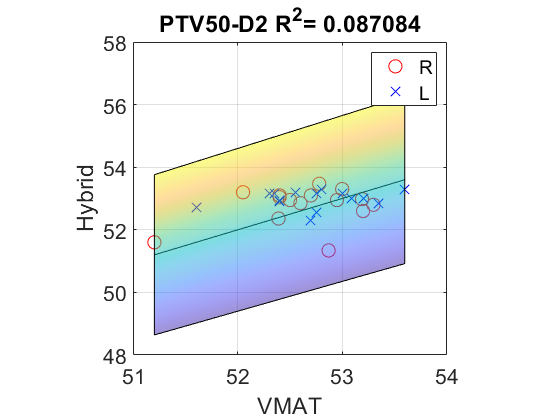}
         \end{subfigure}
           \begin{subfigure}[b]{0.24\textwidth}
         \centering
         \includegraphics[width=\textwidth]{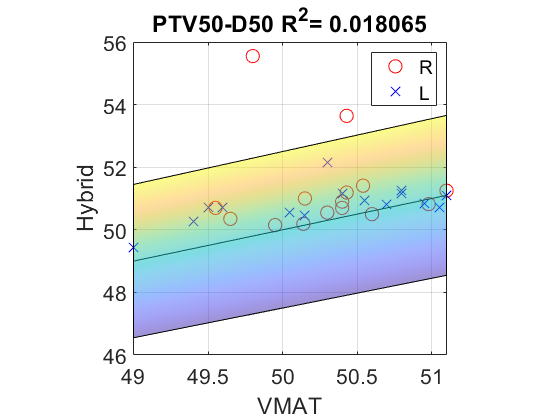}
         \end{subfigure}
           \begin{subfigure}[b]{0.24\textwidth}
         \centering
         \includegraphics[width=\textwidth]{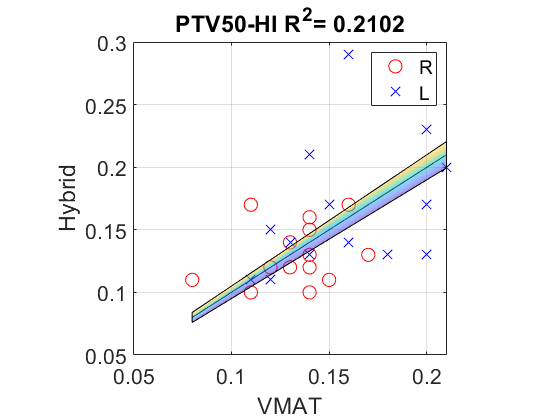}
         \end{subfigure}
         \begin{subfigure}[b]{0.24\textwidth}
         \centering
         \includegraphics[width=\textwidth]{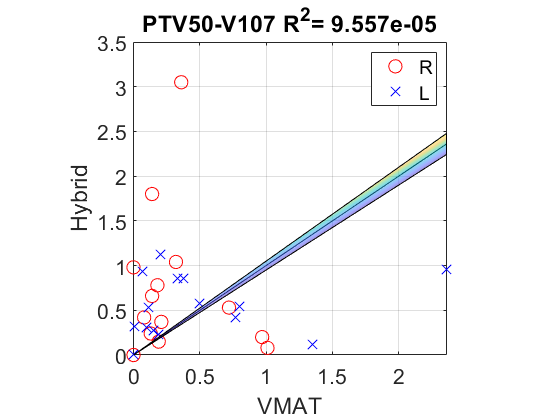}
         \end{subfigure}
           \begin{subfigure}[b]{0.24\textwidth}
         \centering
         \includegraphics[width=\textwidth]{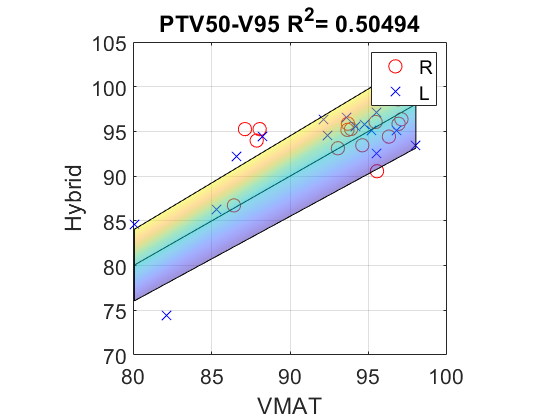}
         \end{subfigure}
           \begin{subfigure}[b]{0.24\textwidth}
         \centering
         \includegraphics[width=\textwidth]{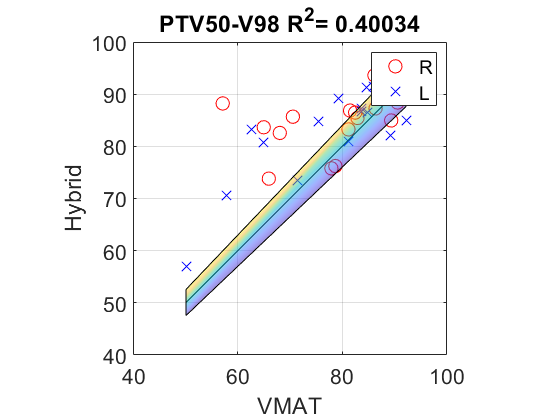}
         \end{subfigure}
           \begin{subfigure}[b]{0.24\textwidth}
         \centering
         \includegraphics[width=\textwidth]{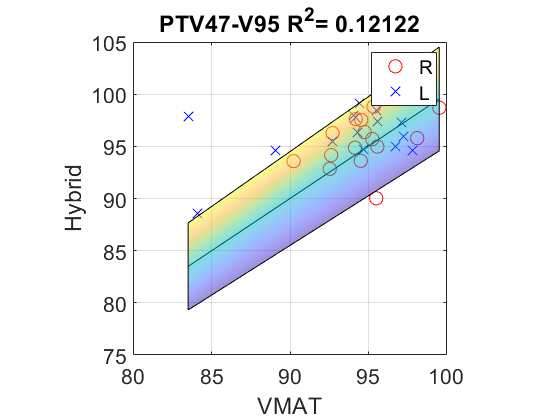}
         \end{subfigure}
         \begin{subfigure}[b]{0.24\textwidth}
         \centering
         \includegraphics[width=\textwidth]{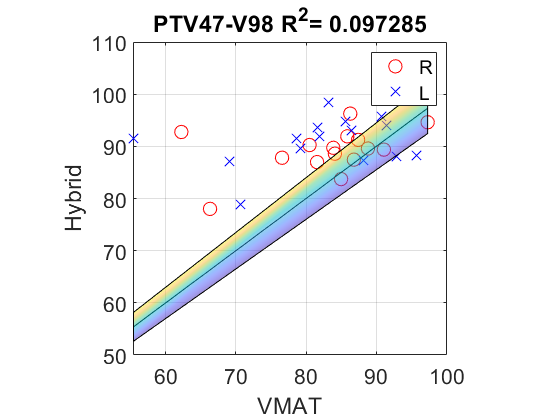}
         \end{subfigure}
\label{fig:determination}
\end{figure}

We have seen previously that VMAT and hybrid share common qualities and shortcomings regarding certain dosimetric parameters. This section will focus on separating these two types of approach proposing a graphical estimation using simple $y=x$ plots (Figure ~\ref{fig:determination}). In addition, we suggest the use of coefficient of determination to refine the interpretation. Points (related to right or left breasts) above the diagonal line ($y=x$) signify a higher value for Hybrid than for VMAT, and vice versa. For points located in the colored band, no conclusion can be drawn as to its significance. For left breasts, V20 of LungIL are often in favour of hybrid, which is not necessarily true for right breasts. However, for V30 and AIV, the result is clear: VMAT puts everyone in agreement offering best results. For the mean dose of heart or dose related to PTV, Hybrid is preferable. If it was chosen to show in this figure only results with an $R^2 < 0.8$ (cases where a majority of points are outside the non-significance area), one parameter seems somewhat disturbing because its distribution of points seems random: LungCL. With an $R^2<0.01$ and a trend (especially for right breasts) in favour of VMAT (although below the clinical goal) few conclusions are possible.

\subsection{Spearman Coefficient Study}
\label{sec:Spearman}
\begin{figure}
 \caption{Spearman correlation analyze of all parameters (Table \ref{tab:param_dos}) and all treatment technics (Table \ref{tab:Beam}). Colored boxes are related to a significant correlations ($\alpha=0.05$)}
 \centerline{\includegraphics[width=24cm]{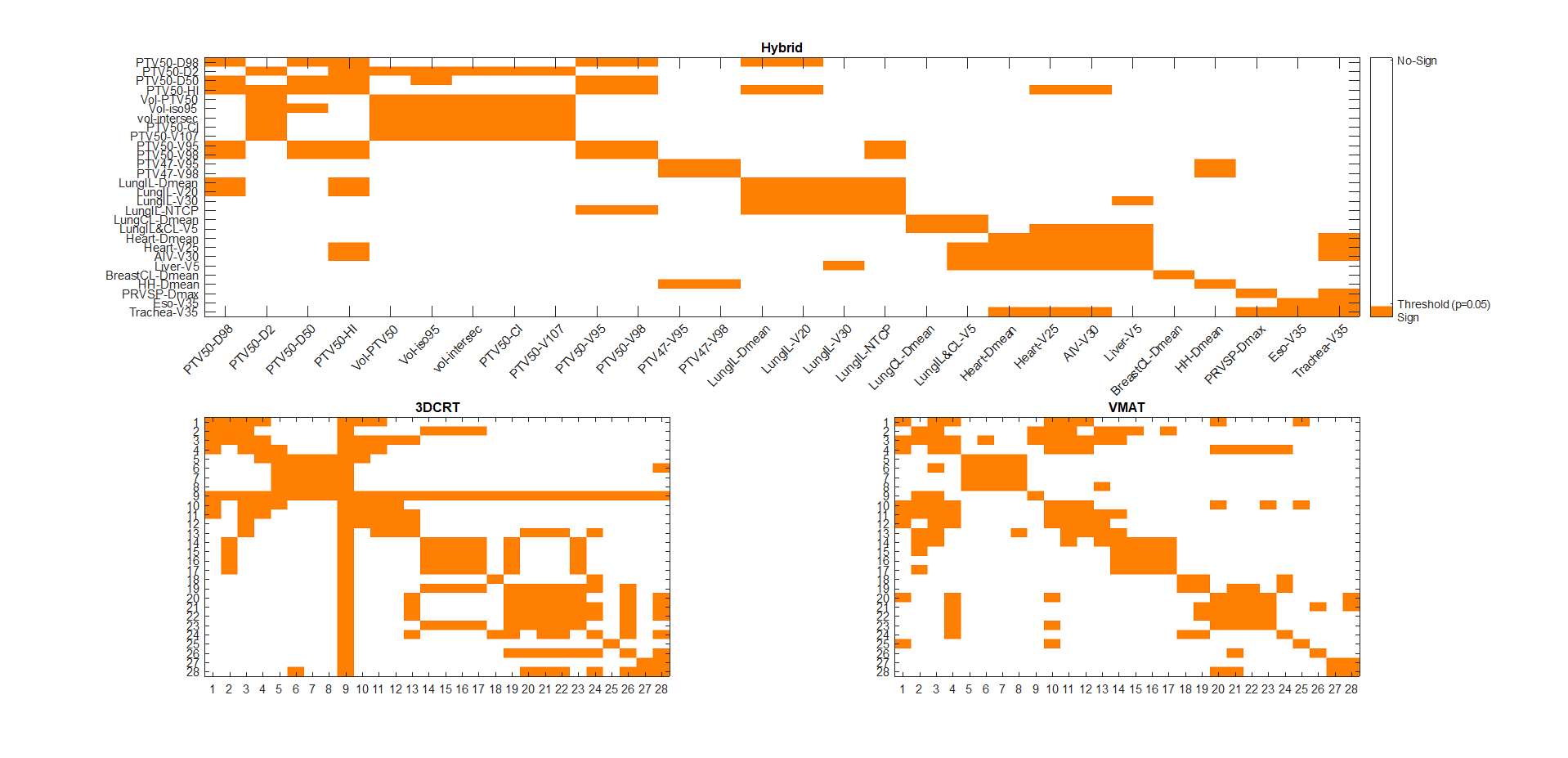}}
 \label{fig:correlation}
\end{figure}

A cross correlation estimation is proposed concerning all available parameters. Figure ~\ref{fig:correlation} summarizes all the results related to the Hybrid, 3DCRT and VMAT plannings. The explanation related to the overall colored line observed with 3DCRT is because le PTV50-V107 is null for all patients. All significant correlations observed close to the diagonal are not surprising (no wonder the V95 and V98 of the target volumes are correlated), what will interest one, are the statistical dependencies outside the diagonals. Thus it is clearly seen that the AIV dose is strongly correlated with HI (for VMAT and Hybrid). This is important because it means to properly preserve the AIV, it is necessary to discover the target volumes. Even more true for the VMAT where HI low values are related to spared liver, heart, HH, etc.. This phenomenon is also observed for the V95 of PTV50. For hybrid, it will mainly be the lungIL dose  which will be correlated to the PTV50 coverage. There are fewer orange boxes for Hybrid than for VMAT outside the diagonal, suggesting that the Hybrid method is more robust and requires less compromise to use it.

\subsection{ROC Study}
\label{sec:ROC}
\begin{figure}
 \caption{ROC Curves concerning AUC>0.7 for R\&L Breasts. Red cross (and value in title of each plot) related to thresholds minimizing sensibility and specificity (=distance to point (0,1)) in order to pass the clinical goal test (Table \ref{tab:param_dos})}
     \centering
     \begin{subfigure}[b]{0.3\textwidth}
         \centering
         \includegraphics[width=\textwidth]{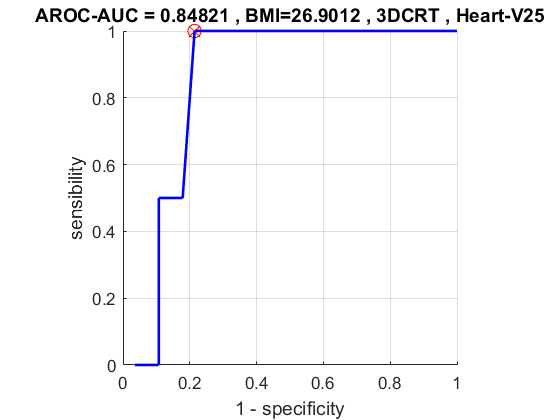}
         \end{subfigure}
           \begin{subfigure}[b]{0.3\textwidth}
         \centering
         \includegraphics[width=\textwidth]{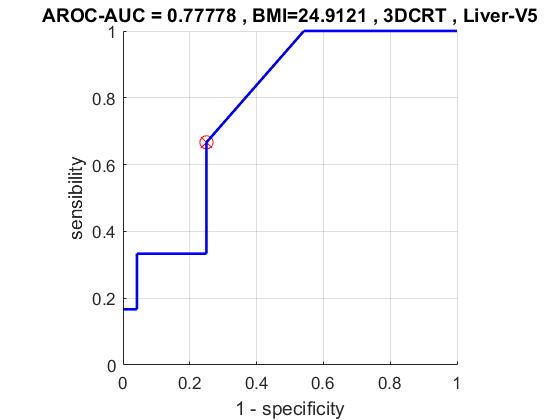}
         \end{subfigure}
           \begin{subfigure}[b]{0.3\textwidth}
         \centering
         \includegraphics[width=\textwidth]{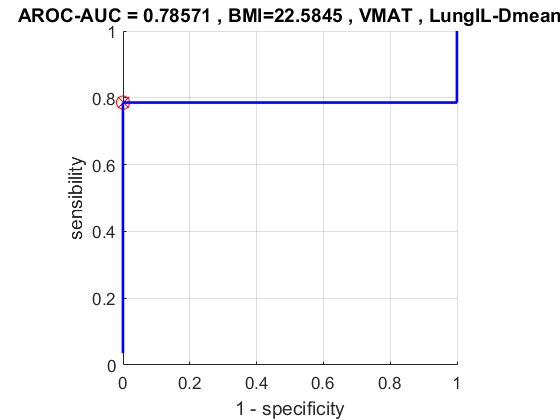}
         \end{subfigure}
         \begin{subfigure}[b]{0.3\textwidth}
         \centering
         \includegraphics[width=\textwidth]{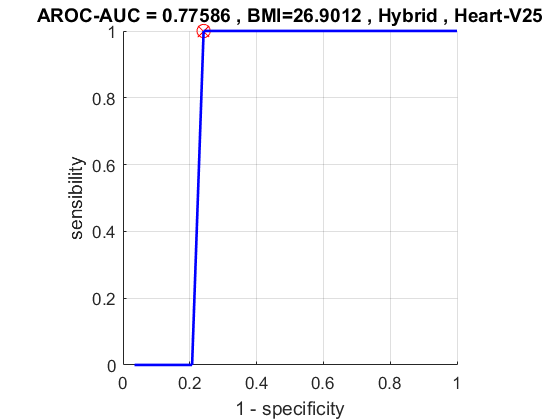}
         \end{subfigure}
           \begin{subfigure}[b]{0.3\textwidth}
         \centering
         \includegraphics[width=\textwidth]{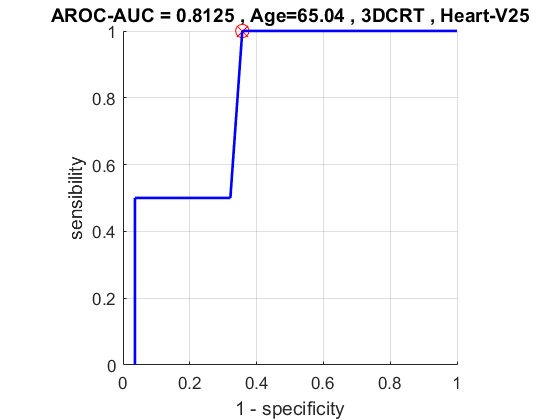}
         \end{subfigure}
           \begin{subfigure}[b]{0.3\textwidth}
         \centering
         \includegraphics[width=\textwidth]{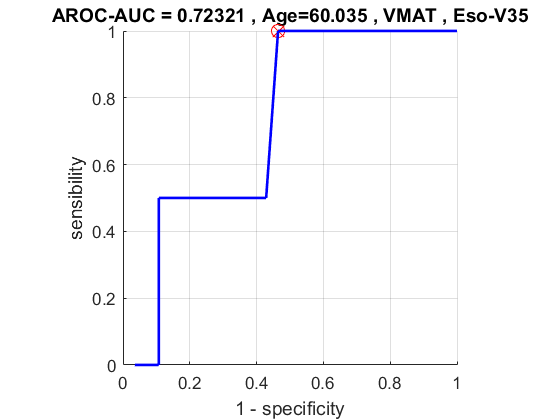}
         \end{subfigure}
         \begin{subfigure}[b]{0.3\textwidth}
         \centering
         \includegraphics[width=\textwidth]{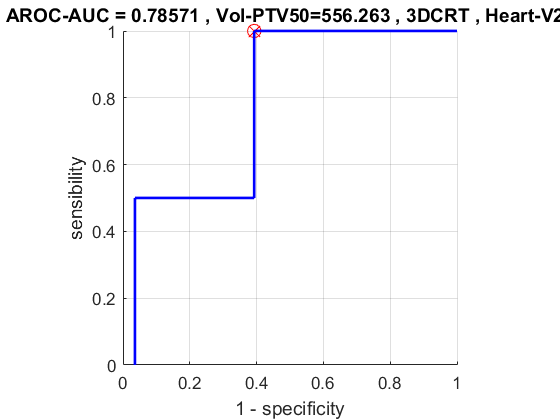}
         \end{subfigure}
         \begin{subfigure}[b]{0.3\textwidth}
         \centering
         \includegraphics[width=\textwidth]{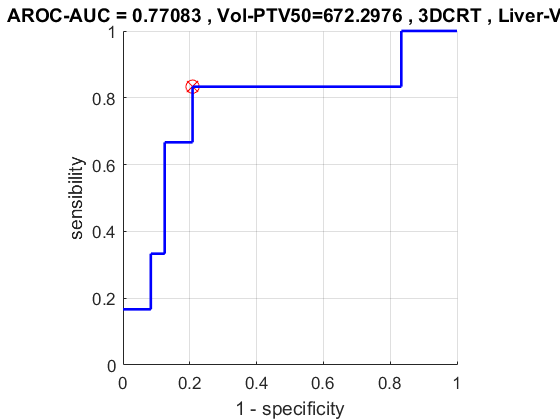}
         \end{subfigure}
           \begin{subfigure}[b]{0.30\textwidth}
         \centering
         \includegraphics[width=\textwidth]{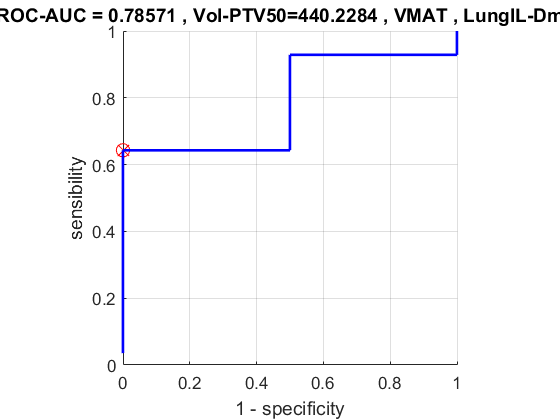}
         \end{subfigure}
\label{fig:rocdg}
\end{figure}

\begin{figure}
 \caption{ROC Curves concerning AUC>0.7 for R Breast. Red cross (and value in title of each plot) related to the threshold minimizing sensibility and specificity (distance to point (0,1)) in order to pass the clinical goal test (Table \ref{tab:param_dos})}
     \centering
        
           \begin{subfigure}[b]{0.3\textwidth}
         \centering
         \includegraphics[width=\textwidth]{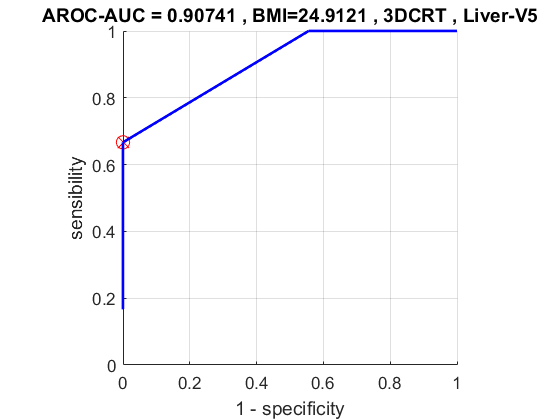}
         \end{subfigure}
         \begin{subfigure}[b]{0.3\textwidth}
         \centering
         \includegraphics[width=\textwidth]{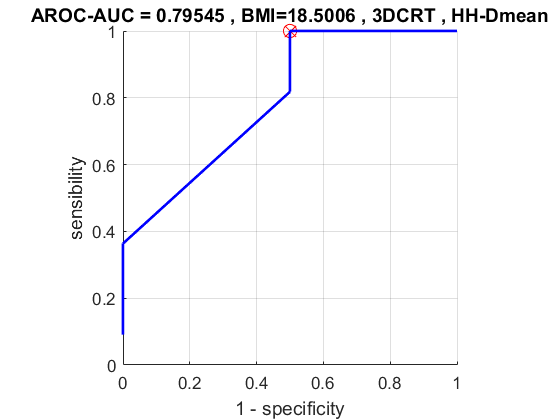}
         \end{subfigure}
         \begin{subfigure}[b]{0.3\textwidth}
         \centering
         \includegraphics[width=\textwidth]{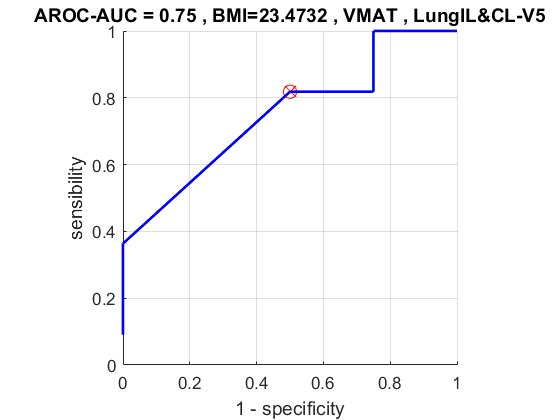}
         \end{subfigure}
           \begin{subfigure}[b]{0.3\textwidth}
         \centering
         \includegraphics[width=\textwidth]{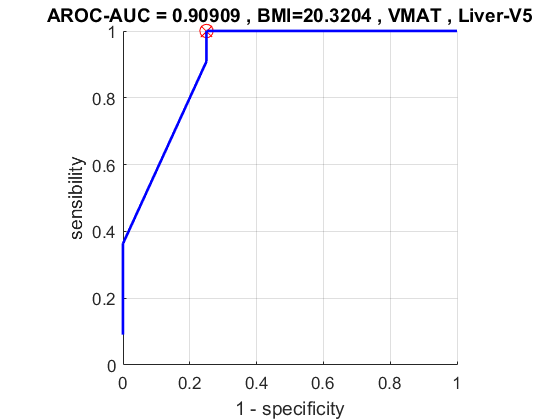}
         \end{subfigure}
           \begin{subfigure}[b]{0.3\textwidth}
         \centering
         \includegraphics[width=\textwidth]{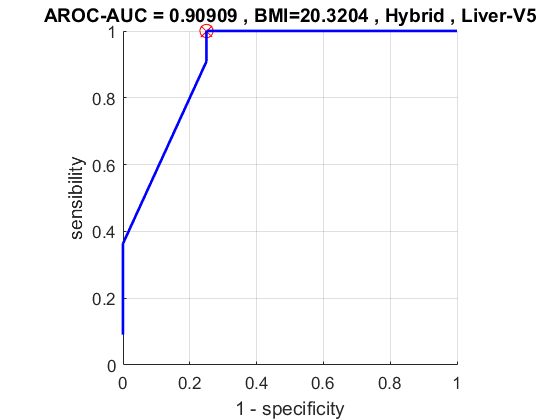}
         \end{subfigure}
         \begin{subfigure}[b]{0.3\textwidth}
         \centering
         \includegraphics[width=\textwidth]{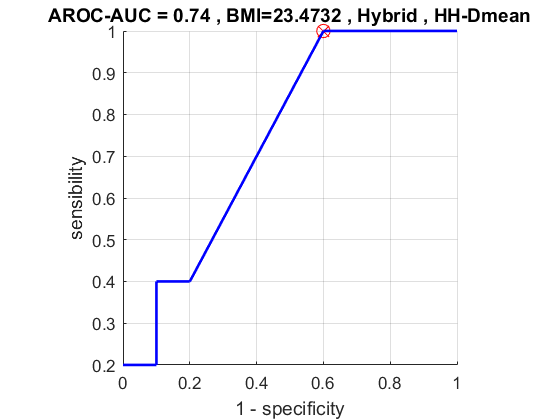}
         \end{subfigure}
         \begin{subfigure}[b]{0.3\textwidth}
         \centering
         \includegraphics[width=\textwidth]{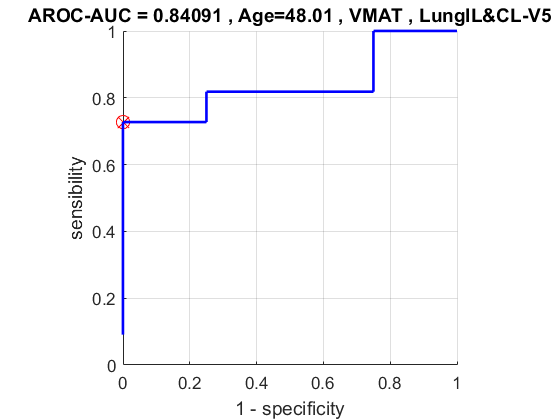}
         \end{subfigure}
           \begin{subfigure}[b]{0.3\textwidth}
         \centering
         \includegraphics[width=\textwidth]{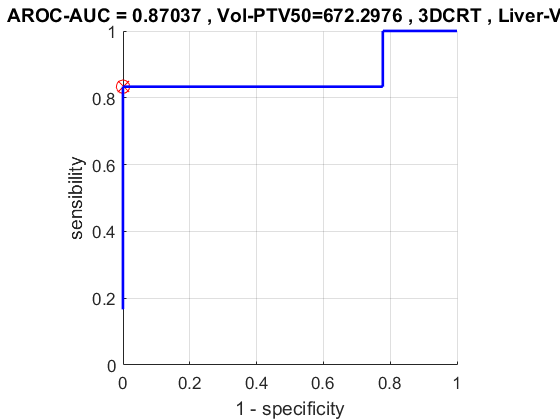}
         \end{subfigure}
           \begin{subfigure}[b]{0.3\textwidth}
         \centering
         \includegraphics[width=\textwidth]{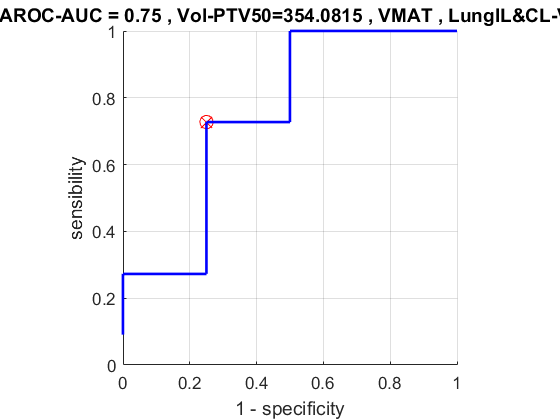}
         \end{subfigure}
         \begin{subfigure}[b]{0.3\textwidth}
         \centering
         \includegraphics[width=\textwidth]{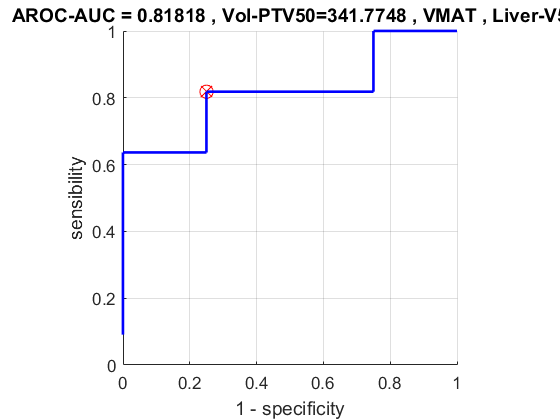}
         \end{subfigure}
         \label{fig:rocd}
\end{figure}

As defined in the Section ~\ref{sec:metrics}, this part is dedicated to the use of ROC curves. Only curves relating to AUC $>0.7$ will be presented so as not to make the reading of this paper cumbersome. This threshold, taken purely arbitrarily, corresponds in some way to a criterion of significance at 70\%. In Figure ~\ref{fig:rocdg} are shown the results concerning both R \& ~L breasts while the Figure ~\ref{fig:rocd} is dedicated to the R breasts. Note that there is not significant ROC curves concerning the L breast. In the first one, we see that age is very little represented. Let's not forget that both sides are considered (right and left), so we will neglect Heart-V25 which is clinically irrelevant for right breasts. This leaves only an interaction between age and esophagus concerning VMAT which is surprising and probably not worth considering.  The BMI and Vol-PTV50 are more relevant because for VMAT, BMI > 22.6 or Vol-PTV50 > 440cc induce a loss of chance to respect the LungIL-Dmean clinical goal (exposed in Table \ref{tab:param_dos}). No such result is obtained for hybrid but it is essential to compare this result with the average obtained in the Table \ref{tab:mean} and which confirms this conclusion. If we focus on R breasts, the results are just as interesting. Indeed, VMAT induces a decrease in the success rate of clinical objectives concerning LungILCL-V5 when BMI > 23.4, age > 48 and Vol-PTV50 > 354. Again, 3DCRT and Hybrid outperform VMAT with a slight advantage for the latter, as it offers much better target volume coverage and therefore allegedly better tumour control. The thresholds established in this study should not be considered as absolute values. Any statistician will recognise statistical instability due to the small number of patients included in the study (15 R and 15 L). This is sufficient to draw conclusions on the trends observed but not to objectify thresholds or create dosimetric references.   

\section{\label{sec:Feasability}Feasibility}

In this section, it is about how the treatments quality control (Delta4) make it possible to validate Hybrid approach generated during this study. It is important to propose a complete study which is not limited to a dosimetric one. The measurement aspect is important because the best theoretical technique could not be considered from a practical point of view, if it does not correspond to technical requirements. In our case, we use the local gamma passing rate criterion which must be $>95\%$ (concerning the 3\%/3mm criterion and the threshold of 10\%) as well as an average gamma lower than 0.4. In order to verify the dosimetric overlap between the two modes of dose delivery (static and arctherapy), we decided to test both arcs and tangentials (with a 9cm cranial offset to test as many points as possible). Results do not take into account the combined results (accumulation of several beams) but only results related to each beam in order to be objective and to limit the compensation effects. The gamma passing rate was equal for the first linac to 97.2\% (std=1.23\%) while for the second one it was 96.4\% (std=1.67\%). The mean gamma was respectively 0.33 and 0.35 and the dose deviation 0.11Gy and 0.16Gy. This is in fact the same order of magnitude of what is observed during VMAT checks. Less than 5\% of dosimetries fail to meet the previously mentioned technical goals and (sometimes) require re-planning.

\section{\label{sec:Conclusion}Conclusion}

The methodologies and conclusions of this study support and improve those established in different studies dealing with hybridization in locoregional breast cancer treatment\cite{LIN2015262,XIE2020e9,LANG2020121}. This study deals with the comparison between 3 types of ballistics used for the treatment of breast cancer. Among them, a hybrid approach mixing the robustness of 3DCRT and the high conformation of VMAT is examined. The idea that has been pointed out for some years concerning the use of arctherapy for such a pathology, lies in the fact of decreasing the low doses deposited in distant healthy tissues. Indeed, in this study, we realize that the doses to the contralateral breast and lung are indeed low (Dmean < 10Gy) but much higher than those observed with 3DCRT. The same is true for the average dose to the heart, which frequently exceeds 5Gy. The strong points of VMAT are the reduction of high doses in organs at risk, so that ipsolateral lung V30, IVA V30 and heart V25 are low with this type of treatment. Concerning D5 of the union of the two lungs or isolateral lung V20, everything suggests that VMAT is less efficient. It has been shown that patients classified as BMI > 23-23.5 or age > 48 or Vol-PTV50 > 350-450cc do not perceive a direct benefit from the use of VMAT due to increased lung doses and failure to achieve certain clinical goals (Tables \ref{tab:param_dos} \& \ref{tab:mean}).
As a result, the hybrid method is positioned (and this was proven in this study) as a robust alternative to VMAT and 3DCRT. For right breasts, it seems clear that this method is preferable, both in terms of lungs, contralateral breast and heart doses. This study reveals, however, that vigilance must be paid to spinal cord, trachea, esophagus and liver. Although the doses are low, it will be appropriate in the context of optimization to add a constraint concerning these organs (without necessarily a large weighting). For the left breast, if the question is more delicate, the answer is just as much so! We would tend to favour the hybrid method, in particular because of the coverage of PTV that it induces, however, VMAT and the reduction of high doses to the heart do not allow us to make a clear decision. In any case, we must not forget that these three methods are more complementary than rivals. Therefore, if the material and human resources allow it, proposing three treatment plans to the physician seems the best solution so that he/she can decide according to what he/she prioritizes. The calculation tools allow to set up these ballistics and associated calculations quickly (< 2h), so it is not a real obstacle. VMAT and hybrid QC show equivalent results in terms of gamma passing rate and mean gamma. If during this study, we have favored parsimony and proposed a hybrid approach with a single arc (in order to minimize the duration of the treatment), it is quite possible (and the dosimetric results are even better while the machine controls are equivalent) to propose a double arc approach (collimator angles set to 0° and 90°) by keeping the limits of arm rotation identical to those presented here. It is likely that the hybrid approach will play an important role in the future of breast cancers treatment, even more so with deep inspiration breath hold technique that move the heart (for left-sided breasts cancer) away from the irradiated area.

\begin{acknowledgments}
We would like to thank M. Brian Baron for his suggestions and for allowing us to benefit from his knowledge and experience in hybrid planning with Pinnacle TPS.

\end{acknowledgments}

\section*{Data Availability Statement}

The data that support the findings of this study are available from the corresponding author upon reasonable request.

\section*{Author Contributions}
\textbf{CV}: Methodology, Investigation, Formal analysis, Writing – original draft, Project administration. \textbf{MP}: Investigation, Conceptualization, Formal analysis, Writing – review \& editing, Project administration.  \textbf{DL}: Investigation,Writing – review \& editing.  \textbf{SP}: Investigation. \textbf{FS}: Investigation, data. \textbf{MAA}: Investigation, Writing – review \& editing.

\section*{Declaration of Competing Interest}
The authors declare that they have no known competing financial interests or personal relationships that could have appeared to influence the work reported in this paper: no conflicts to disclose.




\nocite{*}
\bibliography{aaaBiblio}

\providecommand{\noopsort}[1]{}\providecommand{\singleletter}[1]{#1}%
\begin{thebibliography}{34}%
\makeatletter
\providecommand \@ifxundefined [1]{%
 \@ifx{#1\undefined}
}%
\providecommand \@ifnum [1]{%
 \ifnum #1\expandafter \@firstoftwo
 \else \expandafter \@secondoftwo
 \fi
}%
\providecommand \@ifx [1]{%
 \ifx #1\expandafter \@firstoftwo
 \else \expandafter \@secondoftwo
 \fi
}%
\providecommand \natexlab [1]{#1}%
\providecommand \enquote  [1]{``#1''}%
\providecommand \bibnamefont  [1]{#1}%
\providecommand \bibfnamefont [1]{#1}%
\providecommand \citenamefont [1]{#1}%
\providecommand \href@noop [0]{\@secondoftwo}%
\providecommand \href [0]{\begingroup \@sanitize@url \@href}%
\providecommand \@href[1]{\@@startlink{#1}\@@href}%
\providecommand \@@href[1]{\endgroup#1\@@endlink}%
\providecommand \@sanitize@url [0]{\catcode `\\12\catcode `\$12\catcode
  `\&12\catcode `\#12\catcode `\^12\catcode `\_12\catcode `\%12\relax}%
\providecommand \@@startlink[1]{}%
\providecommand \@@endlink[0]{}%
\providecommand \url  [0]{\begingroup\@sanitize@url \@url }%
\providecommand \@url [1]{\endgroup\@href {#1}{\urlprefix }}%
\providecommand \urlprefix  [0]{URL }%
\providecommand \Eprint [0]{\href }%
\providecommand \doibase [0]{http://dx.doi.org/}%
\providecommand \selectlanguage [0]{\@gobble}%
\providecommand \bibinfo  [0]{\@secondoftwo}%
\providecommand \bibfield  [0]{\@secondoftwo}%
\providecommand \translation [1]{[#1]}%
\providecommand \BibitemOpen [0]{}%
\providecommand \bibitemStop [0]{}%
\providecommand \bibitemNoStop [0]{.\EOS\space}%
\providecommand \EOS [0]{\spacefactor3000\relax}%
\providecommand \BibitemShut  [1]{\csname bibitem#1\endcsname}%
\let\auto@bib@innerbib\@empty
\bibitem [{\citenamefont {Waks}\ and\ \citenamefont
  {Winer}(2019)}]{10.1001/jama.2018.19323}%
  \BibitemOpen
  \bibfield  {author} {\bibinfo {author} {\bibfnamefont {A.~G.}\ \bibnamefont
  {Waks}}\ and\ \bibinfo {author} {\bibfnamefont {E.~P.}\ \bibnamefont
  {Winer}},\ }\bibfield  {title} {\enquote {\bibinfo {title} {{Breast Cancer
  Treatment: A Review}},}\ }\href {\doibase 10.1001/jama.2018.19323} {\bibfield
   {journal} {\bibinfo  {journal} {JAMA}\ }\textbf {\bibinfo {volume} {321}},\
  \bibinfo {pages} {288--300} (\bibinfo {year} {2019})},\ \Eprint
  {http://arxiv.org/abs/https://jamanetwork.com/journals/jama/articlepdf/2721183/jama\_waks\_2019\_rv\_180011.pdf}
  {https://jamanetwork.com/journals/jama/articlepdf/2721183/jama\_waks\_2019\_rv\_180011.pdf}
  \BibitemShut {NoStop}%
\bibitem [{\citenamefont {Franco}\ \emph {et~al.}(2023)\citenamefont {Franco},
  \citenamefont {{De Felice}}, \citenamefont {Jagsi}, \citenamefont {{Nader
  Marta}}, \citenamefont {Kaidar-Person}, \citenamefont {Gabrys}, \citenamefont
  {Kim}, \citenamefont {Ramiah}, \citenamefont {Meattini},\ and\ \citenamefont
  {Poortmans}}]{FRANCO2023100556}%
  \BibitemOpen
  \bibfield  {author} {\bibinfo {author} {\bibfnamefont {P.}~\bibnamefont
  {Franco}}, \bibinfo {author} {\bibfnamefont {F.}~\bibnamefont {{De Felice}}},
  \bibinfo {author} {\bibfnamefont {R.}~\bibnamefont {Jagsi}}, \bibinfo
  {author} {\bibfnamefont {G.}~\bibnamefont {{Nader Marta}}}, \bibinfo {author}
  {\bibfnamefont {O.}~\bibnamefont {Kaidar-Person}}, \bibinfo {author}
  {\bibfnamefont {D.}~\bibnamefont {Gabrys}}, \bibinfo {author} {\bibfnamefont
  {K.}~\bibnamefont {Kim}}, \bibinfo {author} {\bibfnamefont {D.}~\bibnamefont
  {Ramiah}}, \bibinfo {author} {\bibfnamefont {I.}~\bibnamefont {Meattini}}, \
  and\ \bibinfo {author} {\bibfnamefont {P.}~\bibnamefont {Poortmans}},\
  }\bibfield  {title} {\enquote {\bibinfo {title} {Breast cancer radiation
  therapy: A bibliometric analysis of the scientific literature},}\ }\href
  {\doibase https://doi.org/10.1016/j.ctro.2022.11.015} {\bibfield  {journal}
  {\bibinfo  {journal} {Clinical and Translational Radiation Oncology}\
  }\textbf {\bibinfo {volume} {39}},\ \bibinfo {pages} {100556} (\bibinfo
  {year} {2023})}\BibitemShut {NoStop}%
\bibitem [{\citenamefont {Abdollahi}\ \emph {et~al.}(2023)\citenamefont
  {Abdollahi}, \citenamefont {Hadi}, \citenamefont {Mowlavi}, \citenamefont
  {Ceberg}, \citenamefont {Aznar}, \citenamefont {Tabrizi}, \citenamefont
  {Salek}, \citenamefont {Ghodsi},\ and\ \citenamefont
  {Shams}}]{ABDOLLAHI2023100201}%
  \BibitemOpen
  \bibfield  {author} {\bibinfo {author} {\bibfnamefont {S.}~\bibnamefont
  {Abdollahi}}, \bibinfo {author} {\bibfnamefont {M.}~\bibnamefont {Hadi}},
  \bibinfo {author} {\bibfnamefont {A.~A.}\ \bibnamefont {Mowlavi}}, \bibinfo
  {author} {\bibfnamefont {S.}~\bibnamefont {Ceberg}}, \bibinfo {author}
  {\bibfnamefont {M.~C.}\ \bibnamefont {Aznar}}, \bibinfo {author}
  {\bibfnamefont {F.~V.}\ \bibnamefont {Tabrizi}}, \bibinfo {author}
  {\bibfnamefont {R.}~\bibnamefont {Salek}}, \bibinfo {author} {\bibfnamefont
  {A.}~\bibnamefont {Ghodsi}}, \ and\ \bibinfo {author} {\bibfnamefont
  {A.}~\bibnamefont {Shams}},\ }\bibfield  {title} {\enquote {\bibinfo {title}
  {A dose planning study for cardiac and lung dose sparing techniques in left
  breast cancer radiotherapy: Can free breathing helical tomotherapy be
  considered as an alternative for deep inspiration breath hold?}}\ }\href
  {\doibase https://doi.org/10.1016/j.tipsro.2023.100201} {\bibfield  {journal}
  {\bibinfo  {journal} {Technical Innovations \& Patient Support in Radiation
  Oncology}\ }\textbf {\bibinfo {volume} {25}},\ \bibinfo {pages} {100201}
  (\bibinfo {year} {2023})}\BibitemShut {NoStop}%
\bibitem [{\citenamefont {Chung}\ \emph {et~al.}(2013)\citenamefont {Chung},
  \citenamefont {Corbett}, \citenamefont {Moran}, \citenamefont {Griffith},
  \citenamefont {Marsh}, \citenamefont {Feng}, \citenamefont {Jagsi},
  \citenamefont {Kessler}, \citenamefont {Ficaro},\ and\ \citenamefont
  {Pierce}}]{CHUNG2013959}%
  \BibitemOpen
  \bibfield  {author} {\bibinfo {author} {\bibfnamefont {E.}~\bibnamefont
  {Chung}}, \bibinfo {author} {\bibfnamefont {J.~R.}\ \bibnamefont {Corbett}},
  \bibinfo {author} {\bibfnamefont {J.~M.}\ \bibnamefont {Moran}}, \bibinfo
  {author} {\bibfnamefont {K.~A.}\ \bibnamefont {Griffith}}, \bibinfo {author}
  {\bibfnamefont {R.~B.}\ \bibnamefont {Marsh}}, \bibinfo {author}
  {\bibfnamefont {M.}~\bibnamefont {Feng}}, \bibinfo {author} {\bibfnamefont
  {R.}~\bibnamefont {Jagsi}}, \bibinfo {author} {\bibfnamefont {M.~L.}\
  \bibnamefont {Kessler}}, \bibinfo {author} {\bibfnamefont {E.~C.}\
  \bibnamefont {Ficaro}}, \ and\ \bibinfo {author} {\bibfnamefont {L.~J.}\
  \bibnamefont {Pierce}},\ }\bibfield  {title} {\enquote {\bibinfo {title} {Is
  there a dose-response relationship for heart disease with low-dose radiation
  therapy?}}\ }\href {\doibase https://doi.org/10.1016/j.ijrobp.2012.08.002}
  {\bibfield  {journal} {\bibinfo  {journal} {International Journal of
  Radiation Oncology*Biology*Physics}\ }\textbf {\bibinfo {volume} {85}},\
  \bibinfo {pages} {959--964} (\bibinfo {year} {2013})}\BibitemShut {NoStop}%
\bibitem [{\citenamefont {Gleeson}(2022)}]{GLEESON2022264}%
  \BibitemOpen
  \bibfield  {author} {\bibinfo {author} {\bibfnamefont {I.}~\bibnamefont
  {Gleeson}},\ }\bibfield  {title} {\enquote {\bibinfo {title} {Comparing the
  robustness of different skin flash approaches using wide tangents, manual
  flash vmat, and simulated organ motion robust optimization vmat in breast and
  nodal radiotherapy},}\ }\href {\doibase
  https://doi.org/10.1016/j.meddos.2022.04.004} {\bibfield  {journal} {\bibinfo
   {journal} {Medical Dosimetry}\ }\textbf {\bibinfo {volume} {47}},\ \bibinfo
  {pages} {264--272} (\bibinfo {year} {2022})}\BibitemShut {NoStop}%
\bibitem [{\citenamefont {Kwa}\ \emph {et~al.}(1998)\citenamefont {Kwa},
  \citenamefont {Lebesque}, \citenamefont {Theuws}, \citenamefont {Marks},
  \citenamefont {Munley}, \citenamefont {Bentel}, \citenamefont {Oetzel},
  \citenamefont {Spahn}, \citenamefont {Graham}, \citenamefont {Drzymala},
  \citenamefont {Purdy}, \citenamefont {Lichter}, \citenamefont {Martel},\ and\
  \citenamefont {{Ten Haken}}}]{KWA19981}%
  \BibitemOpen
  \bibfield  {author} {\bibinfo {author} {\bibfnamefont {S.~L.}\ \bibnamefont
  {Kwa}}, \bibinfo {author} {\bibfnamefont {J.~V.}\ \bibnamefont {Lebesque}},
  \bibinfo {author} {\bibfnamefont {J.~C.}\ \bibnamefont {Theuws}}, \bibinfo
  {author} {\bibfnamefont {L.~B.}\ \bibnamefont {Marks}}, \bibinfo {author}
  {\bibfnamefont {M.~T.}\ \bibnamefont {Munley}}, \bibinfo {author}
  {\bibfnamefont {G.}~\bibnamefont {Bentel}}, \bibinfo {author} {\bibfnamefont
  {D.}~\bibnamefont {Oetzel}}, \bibinfo {author} {\bibfnamefont
  {U.}~\bibnamefont {Spahn}}, \bibinfo {author} {\bibfnamefont {M.~V.}\
  \bibnamefont {Graham}}, \bibinfo {author} {\bibfnamefont {R.~E.}\
  \bibnamefont {Drzymala}}, \bibinfo {author} {\bibfnamefont {J.~A.}\
  \bibnamefont {Purdy}}, \bibinfo {author} {\bibfnamefont {A.~S.}\ \bibnamefont
  {Lichter}}, \bibinfo {author} {\bibfnamefont {M.~K.}\ \bibnamefont {Martel}},
  \ and\ \bibinfo {author} {\bibfnamefont {R.~K.}\ \bibnamefont {{Ten
  Haken}}},\ }\bibfield  {title} {\enquote {\bibinfo {title} {Radiation
  pneumonitis as a function of mean lung dose: an analysis of pooled data of
  540 patients},}\ }\href {\doibase
  https://doi.org/10.1016/S0360-3016(98)00196-5} {\bibfield  {journal}
  {\bibinfo  {journal} {International Journal of Radiation
  Oncology*Biology*Physics}\ }\textbf {\bibinfo {volume} {42}},\ \bibinfo
  {pages} {1--9} (\bibinfo {year} {1998})}\BibitemShut {NoStop}%
\bibitem [{\citenamefont {White}\ and\ \citenamefont
  {Joiner}(2006)}]{White2006}%
  \BibitemOpen
  \bibfield  {author} {\bibinfo {author} {\bibfnamefont {J.}~\bibnamefont
  {White}}\ and\ \bibinfo {author} {\bibfnamefont {M.~C.}\ \bibnamefont
  {Joiner}},\ }\enquote {\bibinfo {title} {Toxicity from radiation in breast
  cancer},}\ in\ \href {\doibase 10.1007/0-387-25354-8_5} {\emph {\bibinfo
  {booktitle} {Radiation Toxicity: A Practical Guide}}},\ \bibinfo {editor}
  {edited by\ \bibinfo {editor} {\bibfnamefont {W.}~\bibnamefont {Small}}\ and\
  \bibinfo {editor} {\bibfnamefont {G.~E.}\ \bibnamefont {Woloschak}}}\
  (\bibinfo  {publisher} {Springer US},\ \bibinfo {address} {Boston, MA},\
  \bibinfo {year} {2006})\ pp.\ \bibinfo {pages} {65--109}\BibitemShut
  {NoStop}%
\bibitem [{\citenamefont {Cardona-Maya}\ \emph {et~al.}(2023)\citenamefont
  {Cardona-Maya}, \citenamefont {Rojas-López}, \citenamefont {Germanier},
  \citenamefont {Murina},\ and\ \citenamefont {Venencia}}]{cardona2023}%
  \BibitemOpen
  \bibfield  {author} {\bibinfo {author} {\bibfnamefont {A.~M.}\ \bibnamefont
  {Cardona-Maya}}, \bibinfo {author} {\bibfnamefont {J.~A.}\ \bibnamefont
  {Rojas-López}}, \bibinfo {author} {\bibfnamefont {A.}~\bibnamefont
  {Germanier}}, \bibinfo {author} {\bibfnamefont {P.}~\bibnamefont {Murina}}, \
  and\ \bibinfo {author} {\bibfnamefont {D.}~\bibnamefont {Venencia}},\
  }\bibfield  {title} {\enquote {\bibinfo {title} {Experimental determination
  of breast skin dose using volumetric modulated arc therapy and field-in-field
  treatment techniques},}\ }\href {\doibase 10.1017/S1460396922000292}
  {\bibfield  {journal} {\bibinfo  {journal} {Journal of Radiotherapy in
  Practice}\ }\textbf {\bibinfo {volume} {22}},\ \bibinfo {pages} {e59}
  (\bibinfo {year} {2023})}\BibitemShut {NoStop}%
\bibitem [{\citenamefont {Rossi}\ \emph {et~al.}(2021)\citenamefont {Rossi},
  \citenamefont {Virén}, \citenamefont {Heikkilä}, \citenamefont
  {Seppälä},\ and\ \citenamefont {Boman}}]{ROSSI202186}%
  \BibitemOpen
  \bibfield  {author} {\bibinfo {author} {\bibfnamefont {M.}~\bibnamefont
  {Rossi}}, \bibinfo {author} {\bibfnamefont {T.}~\bibnamefont {Virén}},
  \bibinfo {author} {\bibfnamefont {J.}~\bibnamefont {Heikkilä}}, \bibinfo
  {author} {\bibfnamefont {J.}~\bibnamefont {Seppälä}}, \ and\ \bibinfo
  {author} {\bibfnamefont {E.}~\bibnamefont {Boman}},\ }\bibfield  {title}
  {\enquote {\bibinfo {title} {The robustness of vmat radiotherapy for breast
  cancer with tissue deformations},}\ }\href {\doibase
  https://doi.org/10.1016/j.meddos.2020.09.005} {\bibfield  {journal} {\bibinfo
   {journal} {Medical Dosimetry}\ }\textbf {\bibinfo {volume} {46}},\ \bibinfo
  {pages} {86--93} (\bibinfo {year} {2021})}\BibitemShut {NoStop}%
\bibitem [{\citenamefont {Borger}\ \emph {et~al.}(2007)\citenamefont {Borger},
  \citenamefont {Hooning}, \citenamefont {Boersma}, \citenamefont
  {Snijders-Keilholz}, \citenamefont {Aleman}, \citenamefont {Lintzen},
  \citenamefont {{van Brussel}}, \citenamefont {{van der Toorn}}, \citenamefont
  {Alwhouhayb},\ and\ \citenamefont {{van Leeuwen}}}]{BORGER20071131}%
  \BibitemOpen
  \bibfield  {author} {\bibinfo {author} {\bibfnamefont {J.~H.}\ \bibnamefont
  {Borger}}, \bibinfo {author} {\bibfnamefont {M.~J.}\ \bibnamefont {Hooning}},
  \bibinfo {author} {\bibfnamefont {L.~J.}\ \bibnamefont {Boersma}}, \bibinfo
  {author} {\bibfnamefont {A.}~\bibnamefont {Snijders-Keilholz}}, \bibinfo
  {author} {\bibfnamefont {B.~M.}\ \bibnamefont {Aleman}}, \bibinfo {author}
  {\bibfnamefont {E.}~\bibnamefont {Lintzen}}, \bibinfo {author} {\bibfnamefont
  {S.}~\bibnamefont {{van Brussel}}}, \bibinfo {author} {\bibfnamefont {P.-P.}\
  \bibnamefont {{van der Toorn}}}, \bibinfo {author} {\bibfnamefont
  {M.}~\bibnamefont {Alwhouhayb}}, \ and\ \bibinfo {author} {\bibfnamefont
  {F.~E.}\ \bibnamefont {{van Leeuwen}}},\ }\bibfield  {title} {\enquote
  {\bibinfo {title} {Cardiotoxic effects of tangential breast irradiation in
  early breast cancer patients: The role of irradiated heart volume},}\ }\href
  {\doibase https://doi.org/10.1016/j.ijrobp.2007.04.042} {\bibfield  {journal}
  {\bibinfo  {journal} {International Journal of Radiation
  Oncology*Biology*Physics}\ }\textbf {\bibinfo {volume} {69}},\ \bibinfo
  {pages} {1131--1138} (\bibinfo {year} {2007})}\BibitemShut {NoStop}%
\bibitem [{\citenamefont {Taylor}\ \emph {et~al.}(2017)\citenamefont {Taylor},
  \citenamefont {Correa}, \citenamefont {Duane}, \citenamefont {Aznar},
  \citenamefont {Anderson}, \citenamefont {Bergh}, \citenamefont {Dodwell},
  \citenamefont {Ewertz}, \citenamefont {Gray}, \citenamefont {Jagsi},
  \citenamefont {Pierce}, \citenamefont {Pritchard}, \citenamefont {Swain},
  \citenamefont {Wang}, \citenamefont {Wang}, \citenamefont {Whelan},
  \citenamefont {Peto}, \citenamefont {McGale},\ and\ \citenamefont {{Early
  Breast Cancer Trialists’ Collaborative Group}}}]{taylor_estimating_2017}%
  \BibitemOpen
  \bibfield  {author} {\bibinfo {author} {\bibfnamefont {C.}~\bibnamefont
  {Taylor}}, \bibinfo {author} {\bibfnamefont {C.}~\bibnamefont {Correa}},
  \bibinfo {author} {\bibfnamefont {F.~K.}\ \bibnamefont {Duane}}, \bibinfo
  {author} {\bibfnamefont {M.~C.}\ \bibnamefont {Aznar}}, \bibinfo {author}
  {\bibfnamefont {S.~J.}\ \bibnamefont {Anderson}}, \bibinfo {author}
  {\bibfnamefont {J.}~\bibnamefont {Bergh}}, \bibinfo {author} {\bibfnamefont
  {D.}~\bibnamefont {Dodwell}}, \bibinfo {author} {\bibfnamefont
  {M.}~\bibnamefont {Ewertz}}, \bibinfo {author} {\bibfnamefont
  {R.}~\bibnamefont {Gray}}, \bibinfo {author} {\bibfnamefont {R.}~\bibnamefont
  {Jagsi}}, \bibinfo {author} {\bibfnamefont {L.}~\bibnamefont {Pierce}},
  \bibinfo {author} {\bibfnamefont {K.~I.}\ \bibnamefont {Pritchard}}, \bibinfo
  {author} {\bibfnamefont {S.}~\bibnamefont {Swain}}, \bibinfo {author}
  {\bibfnamefont {Z.}~\bibnamefont {Wang}}, \bibinfo {author} {\bibfnamefont
  {Y.}~\bibnamefont {Wang}}, \bibinfo {author} {\bibfnamefont {T.}~\bibnamefont
  {Whelan}}, \bibinfo {author} {\bibfnamefont {R.}~\bibnamefont {Peto}},
  \bibinfo {author} {\bibfnamefont {P.}~\bibnamefont {McGale}}, \ and\ \bibinfo
  {author} {\bibnamefont {{Early Breast Cancer Trialists’ Collaborative
  Group}}},\ }\bibfield  {title} {\enquote {\bibinfo {title} {Estimating the
  {Risks} of {Breast} {Cancer} {Radiotherapy}: {Evidence} {From} {Modern}
  {Radiation} {Doses} to the {Lungs} and {Heart} and {From} {Previous}
  {Randomized} {Trials}},}\ }\href {\doibase 10.1200/JCO.2016.72.0722}
  {\bibfield  {journal} {\bibinfo  {journal} {Journal of Clinical Oncology:
  Official Journal of the American Society of Clinical Oncology}\ }\textbf
  {\bibinfo {volume} {35}},\ \bibinfo {pages} {1641--1649} (\bibinfo {year}
  {2017})}\BibitemShut {NoStop}%
\bibitem [{\citenamefont {Roy}, \citenamefont {Salerno},\ and\ \citenamefont
  {Citrin}(2021)}]{ROY2021155}%
  \BibitemOpen
  \bibfield  {author} {\bibinfo {author} {\bibfnamefont {S.}~\bibnamefont
  {Roy}}, \bibinfo {author} {\bibfnamefont {K.~E.}\ \bibnamefont {Salerno}}, \
  and\ \bibinfo {author} {\bibfnamefont {D.~E.}\ \bibnamefont {Citrin}},\
  }\bibfield  {title} {\enquote {\bibinfo {title} {Biology of radiation-induced
  lung injury},}\ }\href {\doibase
  https://doi.org/10.1016/j.semradonc.2020.11.006} {\bibfield  {journal}
  {\bibinfo  {journal} {Seminars in Radiation Oncology}\ }\textbf {\bibinfo
  {volume} {31}},\ \bibinfo {pages} {155--161} (\bibinfo {year} {2021})},\
  \bibinfo {note} {non-Small Cell Lung Cancer}\BibitemShut {NoStop}%
\bibitem [{\citenamefont {Schröder}\ \emph {et~al.}(2021)\citenamefont
  {Schröder}, \citenamefont {Buchali}, \citenamefont {Windisch}, \citenamefont
  {Vu}, \citenamefont {Basler}, \citenamefont {Zwahlen},\ and\ \citenamefont
  {Förster}}]{cancers13010022}%
  \BibitemOpen
  \bibfield  {author} {\bibinfo {author} {\bibfnamefont {C.}~\bibnamefont
  {Schröder}}, \bibinfo {author} {\bibfnamefont {A.}~\bibnamefont {Buchali}},
  \bibinfo {author} {\bibfnamefont {P.}~\bibnamefont {Windisch}}, \bibinfo
  {author} {\bibfnamefont {E.}~\bibnamefont {Vu}}, \bibinfo {author}
  {\bibfnamefont {L.}~\bibnamefont {Basler}}, \bibinfo {author} {\bibfnamefont
  {D.~R.}\ \bibnamefont {Zwahlen}}, \ and\ \bibinfo {author} {\bibfnamefont
  {R.}~\bibnamefont {Förster}},\ }\bibfield  {title} {\enquote {\bibinfo
  {title} {Impact of low-dose irradiation of the lung and heart on toxicity and
  pulmonary function parameters after thoracic radiotherapy},}\ }\href
  {\doibase 10.3390/cancers13010022} {\bibfield  {journal} {\bibinfo  {journal}
  {Cancers}\ }\textbf {\bibinfo {volume} {13}} (\bibinfo {year} {2021}),\
  10.3390/cancers13010022}\BibitemShut {NoStop}%
\bibitem [{\citenamefont {Chhina}\ \emph {et~al.}(2022)\citenamefont {Chhina},
  \citenamefont {Chau}, \citenamefont {Navarro}, \citenamefont {Sendzik},
  \citenamefont {Frisbee},\ and\ \citenamefont {Frisbee}}]{iiii}%
  \BibitemOpen
  \bibfield  {author} {\bibinfo {author} {\bibfnamefont {J.}~\bibnamefont
  {Chhina}}, \bibinfo {author} {\bibfnamefont {O.~W.}\ \bibnamefont {Chau}},
  \bibinfo {author} {\bibfnamefont {J.}~\bibnamefont {Navarro}}, \bibinfo
  {author} {\bibfnamefont {H.}~\bibnamefont {Sendzik}}, \bibinfo {author}
  {\bibfnamefont {J.}~\bibnamefont {Frisbee}}, \ and\ \bibinfo {author}
  {\bibfnamefont {S.}~\bibnamefont {Frisbee}},\ }\bibfield  {title} {\enquote
  {\bibinfo {title} {The cardiovascular effects of breast cancer radiation
  therapy},}\ }\href {\doibase https://doi.org/10.1096/fasebj.2022.36.S1.R4012}
  {\bibfield  {journal} {\bibinfo  {journal} {The FASEB Journal}\ }\textbf
  {\bibinfo {volume} {36}} (\bibinfo {year} {2022}),\
  https://doi.org/10.1096/fasebj.2022.36.S1.R4012},\ \Eprint
  {http://arxiv.org/abs/https://faseb.onlinelibrary.wiley.com/doi/pdf/10.1096/fasebj.2022.36.S1.R4012}
  {https://faseb.onlinelibrary.wiley.com/doi/pdf/10.1096/fasebj.2022.36.S1.R4012}
  \BibitemShut {NoStop}%
\bibitem [{\citenamefont {Hurkmans}\ \emph {et~al.}(2000)\citenamefont
  {Hurkmans}, \citenamefont {Borger}, \citenamefont {Bos}, \citenamefont {{van
  der Horst}}, \citenamefont {Pieters}, \citenamefont {Lebesque},\ and\
  \citenamefont {Mijnheer}}]{HURKMANS2000145}%
  \BibitemOpen
  \bibfield  {author} {\bibinfo {author} {\bibfnamefont {C.~W.}\ \bibnamefont
  {Hurkmans}}, \bibinfo {author} {\bibfnamefont {J.~H.}\ \bibnamefont
  {Borger}}, \bibinfo {author} {\bibfnamefont {L.~J.}\ \bibnamefont {Bos}},
  \bibinfo {author} {\bibfnamefont {A.}~\bibnamefont {{van der Horst}}},
  \bibinfo {author} {\bibfnamefont {B.~R.}\ \bibnamefont {Pieters}}, \bibinfo
  {author} {\bibfnamefont {J.~V.}\ \bibnamefont {Lebesque}}, \ and\ \bibinfo
  {author} {\bibfnamefont {B.~J.}\ \bibnamefont {Mijnheer}},\ }\bibfield
  {title} {\enquote {\bibinfo {title} {Cardiac and lung complication
  probabilities after breast cancer irradiation},}\ }\href {\doibase
  https://doi.org/10.1016/S0167-8140(00)00152-3} {\bibfield  {journal}
  {\bibinfo  {journal} {Radiotherapy and Oncology}\ }\textbf {\bibinfo {volume}
  {55}},\ \bibinfo {pages} {145--151} (\bibinfo {year} {2000})}\BibitemShut
  {NoStop}%
\bibitem [{\citenamefont {Keffer}, \citenamefont {Guy},\ and\ \citenamefont
  {Weiss}(2020)}]{KEFFER2020238}%
  \BibitemOpen
  \bibfield  {author} {\bibinfo {author} {\bibfnamefont {S.}~\bibnamefont
  {Keffer}}, \bibinfo {author} {\bibfnamefont {C.~L.}\ \bibnamefont {Guy}}, \
  and\ \bibinfo {author} {\bibfnamefont {E.}~\bibnamefont {Weiss}},\ }\bibfield
   {title} {\enquote {\bibinfo {title} {Fatal radiation pneumonitis: Literature
  review and case series},}\ }\href {\doibase
  https://doi.org/10.1016/j.adro.2019.08.010} {\bibfield  {journal} {\bibinfo
  {journal} {Advances in Radiation Oncology}\ }\textbf {\bibinfo {volume}
  {5}},\ \bibinfo {pages} {238--249} (\bibinfo {year} {2020})}\BibitemShut
  {NoStop}%
\bibitem [{\citenamefont {Kang}\ \emph {et~al.}(2021)\citenamefont {Kang},
  \citenamefont {Chen}, \citenamefont {Shi}, \citenamefont {He},\ and\
  \citenamefont {Gao}}]{10.1371/journal.pone.0252552}%
  \BibitemOpen
  \bibfield  {author} {\bibinfo {author} {\bibfnamefont {Z.}~\bibnamefont
  {Kang}}, \bibinfo {author} {\bibfnamefont {S.}~\bibnamefont {Chen}}, \bibinfo
  {author} {\bibfnamefont {L.}~\bibnamefont {Shi}}, \bibinfo {author}
  {\bibfnamefont {Y.}~\bibnamefont {He}}, \ and\ \bibinfo {author}
  {\bibfnamefont {X.}~\bibnamefont {Gao}},\ }\bibfield  {title} {\enquote
  {\bibinfo {title} {Predictors of heart and lung dose in left-sided breast
  cancer treated with vmat relative to 3d-crt: A retrospective study},}\ }\href
  {\doibase 10.1371/journal.pone.0252552} {\bibfield  {journal} {\bibinfo
  {journal} {PLOS ONE}\ }\textbf {\bibinfo {volume} {16}},\ \bibinfo {pages}
  {1--10} (\bibinfo {year} {2021})}\BibitemShut {NoStop}%
\bibitem [{\citenamefont {Liu}\ \emph {et~al.}(2020)\citenamefont {Liu},
  \citenamefont {Chang}, \citenamefont {Lin}, \citenamefont {Lu},\ and\
  \citenamefont {Lai}}]{liu_dosimetric_2020}%
  \BibitemOpen
  \bibfield  {author} {\bibinfo {author} {\bibfnamefont {Y.-C.}\ \bibnamefont
  {Liu}}, \bibinfo {author} {\bibfnamefont {H.-M.}\ \bibnamefont {Chang}},
  \bibinfo {author} {\bibfnamefont {H.-H.}\ \bibnamefont {Lin}}, \bibinfo
  {author} {\bibfnamefont {C.-C.}\ \bibnamefont {Lu}}, \ and\ \bibinfo {author}
  {\bibfnamefont {L.-H.}\ \bibnamefont {Lai}},\ }\bibfield  {title} {\enquote
  {\bibinfo {title} {Dosimetric {Comparison} of {Intensity}-{Modulated}
  {Radiotherapy}, {Volumetric} {Modulated} {Arc} {Therapy} and {Hybrid}
  {Three}-{Dimensional} {Conformal} {Radiotherapy}/{Intensity}-{Modulated}
  {Radiotherapy} {Techniques} for {Right} {Breast} {Cancer}},}\ }\href
  {\doibase 10.3390/jcm9123884} {\bibfield  {journal} {\bibinfo  {journal}
  {Journal of Clinical Medicine}\ }\textbf {\bibinfo {volume} {9}},\ \bibinfo
  {pages} {3884} (\bibinfo {year} {2020})},\ \bibinfo {note} {number: 12
  Publisher: Multidisciplinary Digital Publishing Institute}\BibitemShut
  {NoStop}%
\bibitem [{\citenamefont {Doi}\ \emph {et~al.}(2020)\citenamefont {Doi},
  \citenamefont {Nakao}, \citenamefont {Miura}, \citenamefont {Ozawa},
  \citenamefont {Kenjo},\ and\ \citenamefont {Nagata}}]{doi_hybrid_2020}%
  \BibitemOpen
  \bibfield  {author} {\bibinfo {author} {\bibfnamefont {Y.}~\bibnamefont
  {Doi}}, \bibinfo {author} {\bibfnamefont {M.}~\bibnamefont {Nakao}}, \bibinfo
  {author} {\bibfnamefont {H.}~\bibnamefont {Miura}}, \bibinfo {author}
  {\bibfnamefont {S.}~\bibnamefont {Ozawa}}, \bibinfo {author} {\bibfnamefont
  {M.}~\bibnamefont {Kenjo}}, \ and\ \bibinfo {author} {\bibfnamefont
  {Y.}~\bibnamefont {Nagata}},\ }\bibfield  {title} {\enquote {\bibinfo {title}
  {Hybrid volumetric-modulated arc therapy for postoperative breast cancer
  including regional lymph nodes: the advantage of dosimetric data and safety
  of toxicities},}\ }\href {\doibase 10.1093/jrr/rraa057} {\bibfield  {journal}
  {\bibinfo  {journal} {Journal of Radiation Research}\ }\textbf {\bibinfo
  {volume} {61}},\ \bibinfo {pages} {747--754} (\bibinfo {year}
  {2020})}\BibitemShut {NoStop}%
\bibitem [{\citenamefont {{Marina Hennet}}\ \emph {et~al.}(2022)\citenamefont
  {{Marina Hennet}}, \citenamefont {{Stephan Radonic}}, \citenamefont {{Uwe
  Schneider}},\ and\ \citenamefont {{Matthias
  Hartmann}}}]{marina_hennet_retrospective_2022}%
  \BibitemOpen
  \bibfield  {author} {\bibinfo {author} {\bibnamefont {{Marina Hennet}}},
  \bibinfo {author} {\bibnamefont {{Stephan Radonic}}}, \bibinfo {author}
  {\bibnamefont {{Uwe Schneider}}}, \ and\ \bibinfo {author} {\bibnamefont
  {{Matthias Hartmann}}},\ }\bibfield  {title} {\enquote {\bibinfo {title}
  {Retrospective evaluation of a robust hybrid planning technique established
  for irradiation of breast cancer patients with included mammary internal
  lymph nodes},}\ }\href {\doibase 10.1186/s13014-022-02039-w} {\bibfield
  {journal} {\bibinfo  {journal} {Radiation Oncology}\ }\textbf {\bibinfo
  {volume} {17}},\ \bibinfo {pages} {1--12} (\bibinfo {year} {2022})},\
  \bibinfo {note} {publisher: BMC}\BibitemShut {NoStop}%
\bibitem [{\citenamefont {Ashby}\ and\ \citenamefont
  {Bridge}(2021)}]{ashby_late_2021}%
  \BibitemOpen
  \bibfield  {author} {\bibinfo {author} {\bibfnamefont {O.}~\bibnamefont
  {Ashby}}\ and\ \bibinfo {author} {\bibfnamefont {P.}~\bibnamefont {Bridge}},\
  }\bibfield  {title} {\enquote {\bibinfo {title} {Late effects arising from
  volumetric modulated arc therapy to the breast: {A} systematic review},}\
  }\href {\doibase 10.1016/j.radi.2020.08.003} {\bibfield  {journal} {\bibinfo
  {journal} {Radiography (London, England: 1995)}\ }\textbf {\bibinfo {volume}
  {27}},\ \bibinfo {pages} {650--653} (\bibinfo {year} {2021})}\BibitemShut
  {NoStop}%
\bibitem [{\citenamefont {Cilla}\ \emph {et~al.}(2021)\citenamefont {Cilla},
  \citenamefont {Macchia}, \citenamefont {Romano}, \citenamefont {Morabito},
  \citenamefont {Boccardi}, \citenamefont {Picardi}, \citenamefont {Valentini},
  \citenamefont {Morganti},\ and\ \citenamefont {Deodato}}]{CILLA2021295}%
  \BibitemOpen
  \bibfield  {author} {\bibinfo {author} {\bibfnamefont {S.}~\bibnamefont
  {Cilla}}, \bibinfo {author} {\bibfnamefont {G.}~\bibnamefont {Macchia}},
  \bibinfo {author} {\bibfnamefont {C.}~\bibnamefont {Romano}}, \bibinfo
  {author} {\bibfnamefont {V.~E.}\ \bibnamefont {Morabito}}, \bibinfo {author}
  {\bibfnamefont {M.}~\bibnamefont {Boccardi}}, \bibinfo {author}
  {\bibfnamefont {V.}~\bibnamefont {Picardi}}, \bibinfo {author} {\bibfnamefont
  {V.}~\bibnamefont {Valentini}}, \bibinfo {author} {\bibfnamefont {A.~G.}\
  \bibnamefont {Morganti}}, \ and\ \bibinfo {author} {\bibfnamefont
  {F.}~\bibnamefont {Deodato}},\ }\bibfield  {title} {\enquote {\bibinfo
  {title} {Challenges in lung and heart avoidance for postmastectomy breast
  cancer radiotherapy: Is automated planning the answer?}}\ }\href {\doibase
  https://doi.org/10.1016/j.meddos.2021.03.002} {\bibfield  {journal} {\bibinfo
   {journal} {Medical Dosimetry}\ }\textbf {\bibinfo {volume} {46}},\ \bibinfo
  {pages} {295--303} (\bibinfo {year} {2021})}\BibitemShut {NoStop}%
\bibitem [{\citenamefont {Chen}, \citenamefont {Ramachandran},\ and\
  \citenamefont {Deb}(2020)}]{chen_dosimetric_2020}%
  \BibitemOpen
  \bibfield  {author} {\bibinfo {author} {\bibfnamefont {S.~N.}\ \bibnamefont
  {Chen}}, \bibinfo {author} {\bibfnamefont {P.}~\bibnamefont {Ramachandran}},
  \ and\ \bibinfo {author} {\bibfnamefont {P.}~\bibnamefont {Deb}},\ }\bibfield
   {title} {\enquote {\bibinfo {title} {Dosimetric comparative study of
  {3DCRT}, {IMRT}, {VMAT}, {Ecomp}, and {Hybrid} techniques for breast
  radiation therapy},}\ }\href {\doibase 10.3857/roj.2020.00619} {\bibfield
  {journal} {\bibinfo  {journal} {Radiation Oncology Journal}\ }\textbf
  {\bibinfo {volume} {38}},\ \bibinfo {pages} {270--281} (\bibinfo {year}
  {2020})}\BibitemShut {NoStop}%
\bibitem [{\citenamefont {Veronesi}\ \emph {et~al.}(2008)\citenamefont
  {Veronesi}, \citenamefont {Arnone}, \citenamefont {Veronesi}, \citenamefont
  {Galimberti}, \citenamefont {Luini}, \citenamefont {Rotmensz}, \citenamefont
  {Botteri}, \citenamefont {Ivaldi}, \citenamefont {Leonardi}, \citenamefont
  {Viale}, \citenamefont {Sagona}, \citenamefont {Paganelli}, \citenamefont
  {Panzeri},\ and\ \citenamefont {Orecchia}}]{veronesi_value_2008}%
  \BibitemOpen
  \bibfield  {author} {\bibinfo {author} {\bibfnamefont {U.}~\bibnamefont
  {Veronesi}}, \bibinfo {author} {\bibfnamefont {P.}~\bibnamefont {Arnone}},
  \bibinfo {author} {\bibfnamefont {P.}~\bibnamefont {Veronesi}}, \bibinfo
  {author} {\bibfnamefont {V.}~\bibnamefont {Galimberti}}, \bibinfo {author}
  {\bibfnamefont {A.}~\bibnamefont {Luini}}, \bibinfo {author} {\bibfnamefont
  {N.}~\bibnamefont {Rotmensz}}, \bibinfo {author} {\bibfnamefont
  {E.}~\bibnamefont {Botteri}}, \bibinfo {author} {\bibfnamefont {G.~B.}\
  \bibnamefont {Ivaldi}}, \bibinfo {author} {\bibfnamefont {M.~C.}\
  \bibnamefont {Leonardi}}, \bibinfo {author} {\bibfnamefont {G.}~\bibnamefont
  {Viale}}, \bibinfo {author} {\bibfnamefont {A.}~\bibnamefont {Sagona}},
  \bibinfo {author} {\bibfnamefont {G.}~\bibnamefont {Paganelli}}, \bibinfo
  {author} {\bibfnamefont {R.}~\bibnamefont {Panzeri}}, \ and\ \bibinfo
  {author} {\bibfnamefont {R.}~\bibnamefont {Orecchia}},\ }\bibfield  {title}
  {\enquote {\bibinfo {title} {The value of radiotherapy on metastatic internal
  mammary nodes in breast cancer. {Results} on a large series},}\ }\href
  {\doibase 10.1093/annonc/mdn183} {\bibfield  {journal} {\bibinfo  {journal}
  {Annals of Oncology}\ }\textbf {\bibinfo {volume} {19}},\ \bibinfo {pages}
  {1553--1560} (\bibinfo {year} {2008})}\BibitemShut {NoStop}%
\bibitem [{\citenamefont {Rossi}, \citenamefont {Boman},\ and\ \citenamefont
  {Kapanen}(2019{\natexlab{a}})}]{ROSSI2019117}%
  \BibitemOpen
  \bibfield  {author} {\bibinfo {author} {\bibfnamefont {M.}~\bibnamefont
  {Rossi}}, \bibinfo {author} {\bibfnamefont {E.}~\bibnamefont {Boman}}, \ and\
  \bibinfo {author} {\bibfnamefont {M.}~\bibnamefont {Kapanen}},\ }\bibfield
  {title} {\enquote {\bibinfo {title} {Contralateral tissue sparing in lymph
  node-positive breast cancer radiotherapy with vmat technique},}\ }\href
  {\doibase https://doi.org/10.1016/j.meddos.2018.03.005} {\bibfield  {journal}
  {\bibinfo  {journal} {Medical Dosimetry}\ }\textbf {\bibinfo {volume} {44}},\
  \bibinfo {pages} {117--121} (\bibinfo {year}
  {2019}{\natexlab{a}})}\BibitemShut {NoStop}%
\bibitem [{\citenamefont {{van Duren-Koopman}}\ \emph
  {et~al.}(2018)\citenamefont {{van Duren-Koopman}}, \citenamefont {Tol},
  \citenamefont {Dahele}, \citenamefont {Bucko}, \citenamefont {Meijnen},
  \citenamefont {Slotman},\ and\ \citenamefont
  {Verbakel}}]{VANDURENKOOPMAN2018332}%
  \BibitemOpen
  \bibfield  {author} {\bibinfo {author} {\bibfnamefont {M.~J.}\ \bibnamefont
  {{van Duren-Koopman}}}, \bibinfo {author} {\bibfnamefont {J.~P.}\
  \bibnamefont {Tol}}, \bibinfo {author} {\bibfnamefont {M.}~\bibnamefont
  {Dahele}}, \bibinfo {author} {\bibfnamefont {E.}~\bibnamefont {Bucko}},
  \bibinfo {author} {\bibfnamefont {P.}~\bibnamefont {Meijnen}}, \bibinfo
  {author} {\bibfnamefont {B.~J.}\ \bibnamefont {Slotman}}, \ and\ \bibinfo
  {author} {\bibfnamefont {W.~F.}\ \bibnamefont {Verbakel}},\ }\bibfield
  {title} {\enquote {\bibinfo {title} {Personalized automated treatment
  planning for breast plus locoregional lymph nodes using hybrid rapidarc},}\
  }\href {\doibase https://doi.org/10.1016/j.prro.2018.03.008} {\bibfield
  {journal} {\bibinfo  {journal} {Practical Radiation Oncology}\ }\textbf
  {\bibinfo {volume} {8}},\ \bibinfo {pages} {332--341} (\bibinfo {year}
  {2018})}\BibitemShut {NoStop}%
\bibitem [{\citenamefont {Xu}\ \emph {et~al.}(2019)\citenamefont {Xu},
  \citenamefont {Wang}, \citenamefont {Hu}, \citenamefont {Tian}, \citenamefont
  {Ma}, \citenamefont {Li}, \citenamefont {Dai},\ and\ \citenamefont
  {Wang}}]{xu_locoregional_2019}%
  \BibitemOpen
  \bibfield  {author} {\bibinfo {author} {\bibfnamefont {Y.}~\bibnamefont
  {Xu}}, \bibinfo {author} {\bibfnamefont {J.}~\bibnamefont {Wang}}, \bibinfo
  {author} {\bibfnamefont {Z.}~\bibnamefont {Hu}}, \bibinfo {author}
  {\bibfnamefont {Y.}~\bibnamefont {Tian}}, \bibinfo {author} {\bibfnamefont
  {P.}~\bibnamefont {Ma}}, \bibinfo {author} {\bibfnamefont {S.}~\bibnamefont
  {Li}}, \bibinfo {author} {\bibfnamefont {J.}~\bibnamefont {Dai}}, \ and\
  \bibinfo {author} {\bibfnamefont {S.}~\bibnamefont {Wang}},\ }\bibfield
  {title} {\enquote {\bibinfo {title} {Locoregional irradiation including
  internal mammary nodal region for left-sided breast cancer after breast
  conserving surgery: {Dosimetric} evaluation of 4 techniques},}\ }\href
  {\doibase 10.1016/j.meddos.2018.09.004} {\bibfield  {journal} {\bibinfo
  {journal} {Medical Dosimetry: Official Journal of the American Association of
  Medical Dosimetrists}\ }\textbf {\bibinfo {volume} {44}},\ \bibinfo {pages}
  {e13--e18} (\bibinfo {year} {2019})}\BibitemShut {NoStop}%
\bibitem [{\citenamefont {Rossi}, \citenamefont {Boman},\ and\ \citenamefont
  {Kapanen}(2019{\natexlab{b}})}]{ROSSI2019266}%
  \BibitemOpen
  \bibfield  {author} {\bibinfo {author} {\bibfnamefont {M.}~\bibnamefont
  {Rossi}}, \bibinfo {author} {\bibfnamefont {E.}~\bibnamefont {Boman}}, \ and\
  \bibinfo {author} {\bibfnamefont {M.}~\bibnamefont {Kapanen}},\ }\bibfield
  {title} {\enquote {\bibinfo {title} {Optimal selection of optimization bolus
  thickness in planning of vmat breast radiotherapy treatments},}\ }\href
  {\doibase https://doi.org/10.1016/j.meddos.2018.10.001} {\bibfield  {journal}
  {\bibinfo  {journal} {Medical Dosimetry}\ }\textbf {\bibinfo {volume} {44}},\
  \bibinfo {pages} {266--273} (\bibinfo {year}
  {2019}{\natexlab{b}})}\BibitemShut {NoStop}%
\bibitem [{\citenamefont {Patel}\ \emph {et~al.}(2020)\citenamefont {Patel},
  \citenamefont {Mandal}, \citenamefont {Choudhary}, \citenamefont {Mishra},\
  and\ \citenamefont {Shende}}]{patel_plan_2020}%
  \BibitemOpen
  \bibfield  {author} {\bibinfo {author} {\bibfnamefont {G.}~\bibnamefont
  {Patel}}, \bibinfo {author} {\bibfnamefont {A.}~\bibnamefont {Mandal}},
  \bibinfo {author} {\bibfnamefont {S.}~\bibnamefont {Choudhary}}, \bibinfo
  {author} {\bibfnamefont {R.}~\bibnamefont {Mishra}}, \ and\ \bibinfo {author}
  {\bibfnamefont {R.}~\bibnamefont {Shende}},\ }\bibfield  {title} {\enquote
  {\bibinfo {title} {Plan evaluation indices: {A} journey of evolution},}\
  }\href {\doibase 10.1016/j.rpor.2020.03.002} {\bibfield  {journal} {\bibinfo
  {journal} {Reports of Practical Oncology and Radiotherapy: Journal of
  Greatpoland Cancer Center in Poznan and Polish Society of Radiation
  Oncology}\ }\textbf {\bibinfo {volume} {25}},\ \bibinfo {pages} {336--344}
  (\bibinfo {year} {2020})}\BibitemShut {NoStop}%
\bibitem [{\citenamefont {Tyran}\ \emph {et~al.}(2018)\citenamefont {Tyran},
  \citenamefont {Tallet}, \citenamefont {Resbeut}, \citenamefont {Ferre},
  \citenamefont {Favrel}, \citenamefont {Fau}, \citenamefont {Moureau-Zabotto},
  \citenamefont {Darreon}, \citenamefont {Gonzague}, \citenamefont
  {Benkemouche}, \citenamefont {Varela-Cagetti}, \citenamefont {Salem},
  \citenamefont {Farnault}, \citenamefont {Acquaviva},\ and\ \citenamefont
  {Mailleux}}]{https://doi.org/10.1002/acm2.12398}%
  \BibitemOpen
  \bibfield  {author} {\bibinfo {author} {\bibfnamefont {M.}~\bibnamefont
  {Tyran}}, \bibinfo {author} {\bibfnamefont {A.}~\bibnamefont {Tallet}},
  \bibinfo {author} {\bibfnamefont {M.}~\bibnamefont {Resbeut}}, \bibinfo
  {author} {\bibfnamefont {M.}~\bibnamefont {Ferre}}, \bibinfo {author}
  {\bibfnamefont {V.}~\bibnamefont {Favrel}}, \bibinfo {author} {\bibfnamefont
  {P.}~\bibnamefont {Fau}}, \bibinfo {author} {\bibfnamefont {L.}~\bibnamefont
  {Moureau-Zabotto}}, \bibinfo {author} {\bibfnamefont {J.}~\bibnamefont
  {Darreon}}, \bibinfo {author} {\bibfnamefont {L.}~\bibnamefont {Gonzague}},
  \bibinfo {author} {\bibfnamefont {A.}~\bibnamefont {Benkemouche}}, \bibinfo
  {author} {\bibfnamefont {L.}~\bibnamefont {Varela-Cagetti}}, \bibinfo
  {author} {\bibfnamefont {N.}~\bibnamefont {Salem}}, \bibinfo {author}
  {\bibfnamefont {B.}~\bibnamefont {Farnault}}, \bibinfo {author}
  {\bibfnamefont {M.-A.}\ \bibnamefont {Acquaviva}}, \ and\ \bibinfo {author}
  {\bibfnamefont {H.}~\bibnamefont {Mailleux}},\ }\bibfield  {title} {\enquote
  {\bibinfo {title} {Safety and benefit of using a virtual bolus during
  treatment planning for breast cancer treated with arc therapy},}\ }\href
  {\doibase https://doi.org/10.1002/acm2.12398} {\bibfield  {journal} {\bibinfo
   {journal} {Journal of Applied Clinical Medical Physics}\ }\textbf {\bibinfo
  {volume} {19}},\ \bibinfo {pages} {463--472} (\bibinfo {year} {2018})},\
  \Eprint
  {http://arxiv.org/abs/https://aapm.onlinelibrary.wiley.com/doi/pdf/10.1002/acm2.12398}
  {https://aapm.onlinelibrary.wiley.com/doi/pdf/10.1002/acm2.12398}
  \BibitemShut {NoStop}%
\bibitem [{\citenamefont {Mahé}, \citenamefont {Barillot},\ and\ \citenamefont
  {Chauvet}(2016)}]{MAHE2016S4}%
  \BibitemOpen
  \bibfield  {author} {\bibinfo {author} {\bibfnamefont {M.-A.}\ \bibnamefont
  {Mahé}}, \bibinfo {author} {\bibfnamefont {I.}~\bibnamefont {Barillot}}, \
  and\ \bibinfo {author} {\bibfnamefont {B.}~\bibnamefont {Chauvet}},\
  }\bibfield  {title} {\enquote {\bibinfo {title} {Recommandations en
  radiothérapie externe et curiethérapie (recorad) : 2e édition},}\ }\href
  {\doibase https://doi.org/10.1016/j.canrad.2016.07.014} {\bibfield  {journal}
  {\bibinfo  {journal} {Cancer/Radiothérapie}\ }\textbf {\bibinfo {volume}
  {20}},\ \bibinfo {pages} {S4--S7} (\bibinfo {year} {2016})},\ \bibinfo {note}
  {recorad : Recommandations pour la pratique de la radiothérapie externe et
  de la curiethérapie}\BibitemShut {NoStop}%
\bibitem [{\citenamefont {Lin}\ \emph {et~al.}(2015)\citenamefont {Lin},
  \citenamefont {Yeh}, \citenamefont {Yeh}, \citenamefont {Chang},\ and\
  \citenamefont {Lin}}]{LIN2015262}%
  \BibitemOpen
  \bibfield  {author} {\bibinfo {author} {\bibfnamefont {J.-F.}\ \bibnamefont
  {Lin}}, \bibinfo {author} {\bibfnamefont {D.-C.}\ \bibnamefont {Yeh}},
  \bibinfo {author} {\bibfnamefont {H.-L.}\ \bibnamefont {Yeh}}, \bibinfo
  {author} {\bibfnamefont {C.-F.}\ \bibnamefont {Chang}}, \ and\ \bibinfo
  {author} {\bibfnamefont {J.-C.}\ \bibnamefont {Lin}},\ }\bibfield  {title}
  {\enquote {\bibinfo {title} {Dosimetric comparison of hybrid
  volumetric-modulated arc therapy, volumetric-modulated arc therapy, and
  intensity-modulated radiation therapy for left-sided early breast cancer},}\
  }\href {\doibase https://doi.org/10.1016/j.meddos.2015.05.003} {\bibfield
  {journal} {\bibinfo  {journal} {Medical Dosimetry}\ }\textbf {\bibinfo
  {volume} {40}},\ \bibinfo {pages} {262--267} (\bibinfo {year}
  {2015})}\BibitemShut {NoStop}%
\bibitem [{\citenamefont {Xie}\ \emph {et~al.}(2020)\citenamefont {Xie},
  \citenamefont {Bourgeois}, \citenamefont {Guo},\ and\ \citenamefont
  {Zhang}}]{XIE2020e9}%
  \BibitemOpen
  \bibfield  {author} {\bibinfo {author} {\bibfnamefont {Y.}~\bibnamefont
  {Xie}}, \bibinfo {author} {\bibfnamefont {D.}~\bibnamefont {Bourgeois}},
  \bibinfo {author} {\bibfnamefont {B.}~\bibnamefont {Guo}}, \ and\ \bibinfo
  {author} {\bibfnamefont {R.}~\bibnamefont {Zhang}},\ }\bibfield  {title}
  {\enquote {\bibinfo {title} {Comparison of conventional and advanced
  radiotherapy techniques for left-sided breast cancer after breast conserving
  surgery},}\ }\href {\doibase https://doi.org/10.1016/j.meddos.2020.05.004}
  {\bibfield  {journal} {\bibinfo  {journal} {Medical Dosimetry}\ }\textbf
  {\bibinfo {volume} {45}},\ \bibinfo {pages} {e9--e16} (\bibinfo {year}
  {2020})}\BibitemShut {NoStop}%
\bibitem [{\citenamefont {Lang}\ \emph {et~al.}(2020)\citenamefont {Lang},
  \citenamefont {Loritz}, \citenamefont {Schwartz}, \citenamefont {Hunzeker},
  \citenamefont {Lenards}, \citenamefont {Culp}, \citenamefont {Finley},\ and\
  \citenamefont {Corbin}}]{LANG2020121}%
  \BibitemOpen
  \bibfield  {author} {\bibinfo {author} {\bibfnamefont {K.}~\bibnamefont
  {Lang}}, \bibinfo {author} {\bibfnamefont {B.}~\bibnamefont {Loritz}},
  \bibinfo {author} {\bibfnamefont {A.}~\bibnamefont {Schwartz}}, \bibinfo
  {author} {\bibfnamefont {A.}~\bibnamefont {Hunzeker}}, \bibinfo {author}
  {\bibfnamefont {N.}~\bibnamefont {Lenards}}, \bibinfo {author} {\bibfnamefont
  {L.}~\bibnamefont {Culp}}, \bibinfo {author} {\bibfnamefont {R.}~\bibnamefont
  {Finley}}, \ and\ \bibinfo {author} {\bibfnamefont {K.~S.}\ \bibnamefont
  {Corbin}},\ }\bibfield  {title} {\enquote {\bibinfo {title} {Dosimetric
  comparison between volumetric-modulated arc therapy and a hybrid
  volumetric-modulated arc therapy and segmented field-in-field technique for
  postmastectomy chest wall and regional lymph node irradiation},}\ }\href
  {\doibase https://doi.org/10.1016/j.meddos.2019.08.001} {\bibfield  {journal}
  {\bibinfo  {journal} {Medical Dosimetry}\ }\textbf {\bibinfo {volume} {45}},\
  \bibinfo {pages} {121--127} (\bibinfo {year} {2020})}\BibitemShut {NoStop}%
\end{thebibliography}%

\end{document}